
\input harvmac.tex
\Title{HUTP-92/A065}{\vbox{\centerline{Topological Orbifold Models and}
\vskip2pt\centerline{Quantum Cohomology Rings}}}

\centerline{Eric Zaslow\footnote{$^*$}
{Supported in part by Fannie and John Hertz Foundation.}}

\bigskip\centerline{Lyman Laboratory of Physics}
\centerline{Harvard University}\centerline{Cambridge, MA 02138}

\vskip .3in

We discuss the toplogical sigma model on an orbifold target space.
We describe the
moduli space of classical minima for computing correlation functions
involving twisted operators, and show, through a detailed computation of
an orbifold of ${\bf CP}^1$ by the dihedral group $D_{4},$ how to compute
the complete ring of observables.  Through this procedure, we compute all
the rings from dihedral ${\bf CP}^1$ orbifolds; we note a similarity with
rings derived from perturbed $D-$series superpotentials of the $A-D-E$
classification of $N = 2$ minimal models.  We then consider ${\bf
CP}^2/D_4,$ and show how the
techniques of topological-anti-topological fusion
might be used to compute
twist field correlation
functions for nonabelian orbifolds.

\Date{11/92}

\newsec{Introduction and Summary}
\nref\rDHVW{L. Dixon, J. Harvey, C. Vafa, and E. Witten, Nucl. Phys. {\bf
B261} (1985) 678, and Nucl. Phys. {\bf B274} (1986) 285;
L. Dixon, D. Friedan, E. Martinec, and S. Shenker, Nucl.
Phys. {\bf B282} (1987) 13;
and D. Freed and C. Vafa, Comm. Math Phys. {\bf 110} (1987) 349.}
\nref\rJR{R. Jackiw and C. Rebbi, Phys. Rev. {\bf D13} (1976) 3398; J.
Goldstone and F. Wilczek, Phys. Rev. Lett. {\bf 47} (1981) 986.}
\nref\rWW{X.-G. Wen and E. Witten, Nucl. Phys. {\bf B261} (1985) 651.}
\nref\rCVTAF{S. Cecotti and C. Vafa, Nucl. Phys {\bf B367}
(1991) 3543.}
\nref\rW{E. Witten, ``Mirror Manifolds and Topological Field
Theory," Institute for Advanced Studies preprint IASSNS-HEP 91/83.}
\nref\rLVW{W. Lerche, C. Vafa, and N. Warner, Nucl. Phys. {\bf B324}
(1989) 427.}
\nref\rN{K. S. Narain, Nucl Phys. {\bf B243} (1984) 131.}
\nref\rCHSW{P. Candelas, G. Horowitz, A. Strominger, and E. Witten,
Nucl Phys {\bf B258} (1985) 46.}
\nref\rWSYMM{E. Witten, Nucl Phys {\bf B202} (1982) 253.}
\nref\rRONE{S.-S. Roan, Intl. J. Math. (2) {\bf 1} (1990), 211.}
\nref\rHV{S. Hamidi and C. Vafa, Nucl. Phys. {\bf B279} (1989) 465.}
\nref\rRTWO{S.-S. Roan, Intl. J. Math. (4) {\bf 2} (1991), 439.}
\nref\rBGH{P. Berglund, B. Greene, and T. H\"ubsch,
"Classical $vs.$ Landau-Ginzburg Geometry of Compactification,"
CERN, U.T., and Harvard preprint CERN-TH-6381/92, UTTG-21-91,
HUTMP-91/B315.}
\nref\rGVW{B. R. Greene, C. Vafa, and N. Warner, Nucl. Phys. {\bf B324}
(1989) 371.}
\nref\rC{P. Candelas, ``Lectures on Complex Manifolds," in
{\sl Proceedings of the Winter School on High Energy Physics,}
Puri, India.}
\nref\rWWW{E. Witten, Nucl. Phys. {\bf B258} (1985) 75.}
\nref\rGP{B. R. Greene and M. R. Plesser, Nucl. Phys. {\bf B338} (1990)
15.}
\nref\rIV{K. Intriligator and C. Vafa, Nucl. Phys. {\bf B339} (1990) 95.}
\nref\rVMPL{C. Vafa, Mod. Phys. Lett. {\bf A4} (1989) 1169.}
\nref\rAB{M. F. Atiyah and R. Bott, Ann. Math. {\bf 88} (1968) 451.}
\nref\rCDGP{P. Candelas, X. C. De La Ossa, P. Green, and L Parkes, Nucl.
Phys. {\bf B359} (1991) 21.}
\nref\rCV{S. Cecotti and C. Vafa, ``Massive Orbifolds," SISSA and Harvard
preprint SISSA 44/92/EP, HUTP-92/A013.}
\nref\rM{G. Mackey, Acta Math. {\bf 99} (1958) 265.}
\nref\rFK{H. M. Frakas and I. Kra, {\sl Riemann Surfaces,} Springer,
1980.}
\nref\rDVV{R. Dijkgraaf, H. Verlinde, and E. Verlinde,
Nucl. Phys. {\bf B352} (1991) 59.}
\nref\rFI{P. Fendley and K. Intriligator, Nucl. Phys. {\bf B372} (1992)
533, and Nucl. Phys. {\bf B380} (1992)
265.}
\nref\rCVNEW{S. Cecotti and C. Vafa, ``On Classification of $N = 2$
Supersymmetric Theories,''  Harvard and SISSA preprints HUTP-92/A064 and
SISSA-203/92/EP.}
\nref\rGATO{B. Gato, Nucl. Phys. {\bf B334} (1990) 414.}
\nref\rZ{A. B. Zamolodchikov, JETP Lett. {\bf 43} (1986) 731.}
\nref\rCVER{S. Cecotti and C. Vafa, Phys. Rev. Lett. {\bf 68}
(1992) 903.}
\nref\rPIII{A. R. Its and V. Yu. Novokshenov, {\sl The Isomonodromic
Deformation Method in the Theory of Painlev\'e Equations,} Lecture Notes
in Mathematics 1191, Springer-Verlag, New York 1986; B.M. McCoy, C.A.
Tracy, and T.T. Wu, J. Math Phys. {\bf 18} (1977) 1058.}
\nref\rBD{A. V. Kitaev, ``The Method of Isomonodromy Deformations for
Degenerate Third Painlev\'e Equation," in {\sl Questions of Quantum Field
Theory and Statistical Physics 8,} (Zap. Nauch. Semin. LOMI v. 161), V. N.
Popov and P. P. Kulish, eds, Nauka (Leningrad).}
\nref\rWLH{W. A. Leaf-Herrmann, Harvard preprint HUTP-91/AO61.}

Orbifolds define consistent string vacua \rDHVW.  Therefore, we may wonder
whether the string theories described by orbifolds have a simple
topological description, or we may inquire about topological properties -
for example Yukawa couplings of fermion generations - of string theories
with orbifold compactifications.  Such knowledge can also be applied to
the non-topological theory as well.
We consider topological sigma models on
orbifolds of Kahler manifolds.  These theories are defined by twisting
the $N = 2$ supersymmetric sigma models, and have associated with them
a ring of observables.  This ``quantum ring" is a generalization of the
chiral primary ring to models which are not conformal field
theories. The discussion of these rings - their characterization and
product structure - for topological
orbifold models is the focus of this paper.

The observables of the (untwisted) topological sigma model are described
by cohomology classes of the target space.  Interactions are treated by
taking intersections of homology cycles in the moduli space of
holomorphic maps (section two).  An orbifold is a possibly singular space,
defined by equating points related by the group action.
In order for the {\sl orbifold} to have a sigma model description at the
non-singular points, the metric and complex structure must be preserved
by the action of the group on the target space.  We show (section three)
that
the observables of the orbifold model are described in terms of the
cohomology of the fixed point manifolds of the group elements.
At the singular points
of the group action, there is an identification of tangent space vectors.
Since the fermions of the sigma model have tangent space indeces, the
fermionic sector of a twisted state obeys twisted boundary conditions.
These conditions lead quite generally \rJR \rWW\ to a fractional fermion number
assigned to the vacuum in this sector.  Thus twisted states have a
shifted fermion number.  By analogy with the familiar correspondence
between topological observables and cohomology elements (for untwisted
theories), we may assign Hodge labels based on the chiral fermion numbers
of observables.  In this way, we describe the ``cohomology'' of the
singular orbifold.  We show that Poincar\'e duality is preserved, and
in the case of a Calabi-Yau orbifold by
a group action which preserves the unique $(d,0)$ form, this
``cohomology'' has the Hodge diamond we would expect from a
Calabi-Yau manifold.
In fact, in several examples (section four) we show that this cohomology
is precisely that of the manifold one gets by resolution of the
singularities.  Another check is agreement with the
appropriate Landau-Ginzburg orbifold theory,
when the manifold in question is a Calabi-Yau variety
defined by a quasi-homogeneous
polynomial.  We offer no general proof of this equivalence.

Computation of the product structure of the ring of observables involves
intersection numbers in an appropriate moduli space.  For a correlation
function involving several $g_i-$twisted observables inserted at points
$p_i$ on a Riemann surface, the moduli space is holomorphic maps having
proper monodromies around these points, or equivalently, holomorphic
equivariant maps from an appropriate branched cover of the Riemann
surface.  We use this formalism in computing an explicit example - a
detailed computation of the complete chiral ring for the orbifold of
${\bf CP}^1$ by the dihedral group $D_4$ (section five).  These findings
can be the generalized to the higher even dihedral groups $D_{2k}$ and
odd groups $D_{2k+1}.$
(section six) or to a higher dimensional target space (section seven).
With knowledge of these rings, and in particular behavior
under scale transformations, we can use recent techniques \rCVTAF\ to try
to compute the proper normalization of twist operators in the
conformal limit of large radius (section eight).
The ${\bf CP}^1$ orbifolds reduce to abelian orbifolds in this limit, and
the requirement of regularity fixes the boundary conditions, giving the
twist field correlations.  For higher dimensional spaces, it is unclear
whether regularity is enough to determine the solution.

\newsec{Topological Sigma Models and Quantum Rings}

Let us briefly recall
the topological sigma model on a Kahler manifold, $K.$  In this case, the
action can be derived as a twisted $N = 2$ model.  This twisting leads to
an isomorphism (as vector spaces) between local BRST observables and the
states of the chiral-primary ring.  Specifically, we have \rW
\eqn\action{
S = 2t\int_\Sigma \!d^2\!z\, {1\over 2} g_{IJ}
\partial_z\phi^I\partial_{\overline{z}} \phi^J +
i\psi_{-}^{\overline{i}} D_{z} \psi_-^i g_{\overline{i} i} +
i\psi_+^{\overline{i}} D_{\overline{z}} \psi_+^i g_{\overline{i} i}
+ R_{i\overline{i}j\overline{j}} \psi_+^i \psi_+^{\overline{i}}
 \psi_-^j \psi_-^{\overline{j}}
.}
Here $\Sigma$ represents the Riemann surface, which, for our purpose will
always be of genus zero, $g_{IJ}$ and $R_{i\overline{i}j\overline{j}}$
are respectively the metric and Riemann tensor of the target space.  $D$
is the pull-back onto $\Sigma$ of the connection under the
map, $\Phi.$  The $N = 2$ structure implies a holomorphic $U(1)$
current, by
which we may twist the energy-momentum tensor.  Mathematically, this is
equivalent to redefining the bundles in which the fields live.
Specifically, we now take $\psi_+^i \in \Phi^{*}(T^{1,0})$ and
$\psi_-^{\overline{i}} \in \Phi^{*}(T^{0,1}),$
and we put $\psi_+^{\overline{i}} \in \Omega^{1,0}
(\Sigma ; \Phi^{*}(T^{0,1})),$ and
$\psi_-^i \in
\Omega^{0,1}(\Sigma;\Phi^{*}(T^{1,0})),$ that is, they combine to form
a one-form on $\Sigma$
with values in the pull-back of the tangent space of $K:$ call
these components $\psi_z^{\overline{i}}$ and $\psi_{\overline{z}}^i$
respectively.
These redefinitions correspond to shifting the spins of the fields by
\eqn\twista{\eqalign{& T \rightarrow T - {1\over2}\partial J \cr
& \overline{T} \rightarrow \overline{T} + {1\over2}\overline{\partial}
\, \overline{J} }}
which is equivalent to adding a background gauge field to the
spin connection.
To make this theory topological, we reinterpret the supersymmetry
transformation as a BRST transformation associated to a topological
symmetry (in order for this to close off-shell, more fields must be
introduced) \rW.  We make the replacement
\eqn\twist{ Q_L + Q_R \rightarrow Q_{BRST}.}
Thus the topological observables are precisely the chiral-chiral fields,
and when the original model in a conformal field theory, i.e. when $K$
is a Calabi-Yau manifold, the elements of the BRST cohomology correspond
precisely with the chiral-primary ring of the conformal theory \rLVW.
When the manifold is
not Calabi-Yau, the topological theory is still well-defined, and the
ring of observables generalizes the chiral primary ring, and can be
thought of as a ``quantum cohomology ring."

[Note that there is another ``twist" we may perform, which, due to a
global anomaly, is only defined on a manifold with vanishing first Chern
class, i.e. a Calabi-Yau manifold.  The observables in this theory have a
different cohomological description \rW.]

We have
\eqn\topact{
S = it\int_\Sigma \!d^2\!z\,\lbrace Q,V \rbrace + t\int_\Sigma
\Phi^{*}(k),}
where $V$ is an
appropriate pre-potential (see \rW)
and $\Phi^{*}(k)$ is the pull-back of the Kahler form and.
The second term in \topact\
a topological term, and for the moment, we restrict ourselves to maps
$\Phi$ within a given component of the space of maps.  That is,
we take maps
of a given instanton number, so that the second term in \topact\ is
constant in this component of maps from $\Sigma$ to $K$.
By standard arguments based on the
vanishing of all correlation functions with Q-exact terms, our
calculations reduce to a semi-classical treatment.  That is, we may
take the large $t$ limit, and restrict ourselves to the moduli
space of classical minima, which occur when
\eqn\holo{\partial_{\overline{z}} \phi^i =
\partial_{z} \phi^{\overline{i}} = 0,}
i.e. $\Phi$ is a holomorphic map (this prescription for forming the
topological model is therefore dependent on the complex structure).
Thus the moduli space for this problem is
\eqn\modsp{\hbox{$\cal M$} \equiv \lbrace \Phi : \Sigma
\rightarrow K | \Phi \hbox{ holomorphic} \rbrace
= \bigoplus_i {\hbox{$\cal M$}}_i,}
where $i$ labels the instanton number.

The correspondence between the
cohomology of the target space and the local observables (BRST
cohomology) is described by replacing form components by the fermion
fields.  Let $A = A_{I \overline{J}}dz^{I} d{\overline z}^{\overline
J}$ be a form on $K$, written in local coordinates, where $I$ and $J$
are multi-indices.  The corresponding operator, ${\hbox{$\cal
O$}}_A,$ is obtained by replacing $dz \rightarrow
\psi_z^{\overline i}$ and $d\overline{z} \rightarrow
\psi_{\overline{z}}^i.$
The isomorphism of cohomologies is described by
the equation
\eqn\isomo{\lbrace Q,{\hbox{$\cal O$}}_A \rbrace = -{\hbox{$\cal
O$}}_{dA}.}
Since BRST trivial operators annihilate all correlators, expectation
values only depend on BRST cohomology classes.  If we label observables
by their corresponding forms, this means we may choose the forms to have
delta-function support on the manifold to which they are Poincar\'e-dual.
This way of representing the observables clarifies the
degree zero instanton sector contribution to observables:  the
correlation functions will have non-zero contribution only at the
points of
intersections of the representative manifolds.  Now the degree
zero holomorphic maps are simply constants, so the integral over
${\hbox{$\cal M$}}_0$
is just an integral over $K.$  Thus, because of the
cancellation of bosonic and fermionic determinants familiar to
topological theories, the degree zero correlations are precisely the
intersection numbers of the forms representing the observables.

Generally, the correlation function must be evaluated by considering the
contribution from each componenent of moduli space.  This is done as
follows \rW.  At a given component of moduli space ${\hbox{$\cal M$}}_i,$
we define a manifold
$L_j \subset {\hbox{$\cal M$}}_i$ for each
observable ${\hbox{$\cal O$}}_{j}(x_{j})$ to be the set of maps
in ${\hbox{$\cal M$}}_i$
which take $x_j$ to a point in the manifold representing the
form corresponding to ${\hbox{$\cal O$}}_{j}.$  Then the $i^{th}$ sector
contribution to the correlation of any number of observables is given by
the intersection number of the $L_j.$  This is equivalent to integrating
over the pullbacks of the forms by the evaluation maps at the points of
insertions.  In equations:
\eqn\cool{\langle \prod_{j=1}^{n}{\cal O}_{A_j}(p_j) \rangle =
\bigcap_{j=1}^{n}\left({\hbox{ev}_j^{-1}L_j}\right) =
\sum_i \int_{{\cal M}_i}\prod_{j=1}^{n}\left({\hbox{ev}_j^*A_j}\right),}
where the evaluation map $\hbox{ev}_j: {\cal M} \rightarrow K$ is defined
by $\hbox{ev}_j(\Phi) = \Phi(p_j).$  Here we have ignored the second
term in \topact.  This is a topological term which has a constant
value in each component of moduli space.  Thus, if $S_i$ represents
the value in the $i-$th component of moduli space, then the $i-$th
term in \cool\ must be weighted by ${\hbox{e}}^{-S_i}.$
Note that the moduli space may need
to be compactified in order to have a sensible intersection theory.

It is instructive for us to discuss the ${\bf CP}^n$ model as an
example \rN.
Here we have $K = {\bf CP}^n,$ which has $h_{ii} = 1, i = 0...n,$ with all
other Hodge numbers vanishing.  The intersection theory of nontrivial
cycles is very simple, then.  The intersection number of homology cycles
is one if the codimensions sum to $n,$ zero otherwise.  Basically, this
is because $L_i$ of codimension $k$ can be taken to be the ${\bf
CP}^{n-k}$ defined by setting $k$ coordinates equal to zero, in an
appropriate basis.

Consider ${\cal M}_k,$ i.e. the holomorphic
maps of degree $k$ from ${\bf CP}^1$
to ${\bf CP}^n$ (we consider genus zero correlations, for these define
the ring).\footnote{$^1$}{I thank S. Axelrod for explaining this to me.}
These are described by $(n+1)-$tuples of homogeneous
polynomials of degree $k$ in two variables, which act as shown below:
\eqn\modcpn{\Phi(X,Y) = \left(\matrix{\phi_{00}&...&\phi_{0k}\cr
\phi_{10}&...&\phi_{1k}\cr
.&.&.\cr
.&.&.\cr
\phi_{n0}&...&\phi_{nk}}\right)
\left(\matrix{X^k\cr
X^{k-1}Y\cr
.\cr
.\cr
Y^k}\right).}
The homogeneity property insures that scale changes on $(X,Y),$ which are
trivial on the ${\bf CP}^1,$ only result in scale changes on the
$(n+1)-$tuple.  Now we should ensure that the polynomials, defined by the
matrix rows, do not have common zeros.  This would make the map $\Phi$
ill-defined.  However, in the $compactified$ moduli space, we allow such
points, which can occur as limits of well-defined maps.  Basically, if
there is a common root, we can factor it out of the $(n+1)-$tuple and get
a new, well-defined map (of a lower degree).  Thus the only requirement
we make on the matrix elements $\phi_{ij}$ is that they are not all zero.
Of course, the matrix $\Phi$ is only defined modulo an overall scale.
So we have shown
\eqn\modspcpn{{\cal M}_k \cong {\bf CP}^{(n+1)(k+1)-1}.}

The cohomology ring of $K$ has a single generator $X$ with $X^{n+1} = 0.$
The quantum ring is defined by the correlation functions.  Consider the
correlator $\langle X^aX^bX^c \rangle.$  This will be nonzero if there is
a $k$ such that $(n+1)(k+1)-1 = a + b + c.$  In this case, the instanton
action is $\e{-kA} \equiv \beta^k,$ where $A$ is the one-instanton
action.  All these correlators derive simply from the relation
\eqn\cpnring{X^{n+1} = \beta,}
which defines the chiral ring.  Note that the chiral fermion number is
conserved if we make the artificial assignment of $n+1$ to the chiral
fermion number of $\beta.$

\newsec{The Orbifold Theory}

We would like to study these theories when the target space is an
orbifold, i.e. we consider the quotient $K/G$ of a Kahler manifold
under a group $G$, which acts on this manifold by isometry.
Thus, the metric will be well-defined on the quotient space - the inner
product of two vectors in $K/G$ may be computed by choosing any lift
of the vectors to $K$ and using the metric on $K;$ $G$-invariance
guarantees independence of the particular lift.
Furthermore, we will assume
that the action of $G$ preserves the complex structure.  That is,
\eqn\comp{g_* \circ J = J \circ g_*
\hbox{ for all } g \,\, \epsilon \,\, G,}
where the asterisk
represents push-forward of vectors, and $J$ is the complex structure.
When $G$ acts with fixed points,
the orbifold will have a set of singular points, though the
string theory is not necessarily singular.
If the manifold is not
Calabi-Yau, then the quantum field theory is not conformal and not a
string vacuum \rCHSW; for $K/G$ to be a ``Calabi-Yau orbifold",
we must have
that $G$ leaves invariant the unique holomorphic $(d,0)$ form
under pull-back.
In either case, though, the topological sigma model is
well-defined.\footnote{$^1$}{As stated in \rW,
this follows from the positivity of the fermionic determinant, which
allows us to define it as a function of the moduli.  In
general the fermionic determinant gives a line bundle over the moduli
space of theories, which will lead to an anomaly.  The anomaly
cancellation condition for the topological theory of the inequivalent
twist is that the manifold be Calabi-Yau.}

To properly consider the full orbifold theory, we must
specify the action of the group $G$ on every operator in the theory.  In
particular, if we were considering the orbifold as a space for string
compactification, then we would need to specify the action of $G$ on the
fields representing the rank $16$ gauge group.  This can lead to
phenomenologically desirable symmetry breaking.  In any case,
we see that the proper definition
of a $g$-twisted state, $\Theta$ (where theta may be any type of quantum
field), is that
\eqn\gtwist{\Theta(\sigma + 2\pi) = g \circ \Theta(\sigma),}
where $\sigma$ is the coordinate along the string.

For our purposes, we will
restrict attention to the twisted $N = 2$ theory at hand.
The action of $G$
on the bosonic fields $\Phi$ is the action considering the fields as
coordinates on the manifold $K.$  On the fermionic fields, which involve
(pull-backs of) the tangent bundle $TK,$ the action is induced from the
coordinate action by the push-forward of vectors.  At any point $p$ on
$K,$ the tangent space $T_p$ is identified with $T_{gp}.$  However, at
a fixed point $f$ of $g$ (i.e. $gf = f$), we must identify tangent
vectors in $T_f$ related by the action of $g.$
More precisely, we must identify all tangent vectors related by the
stabilizer group of elements fixing $f$:  $S(f) =
\lbrace g \,\, \epsilon \,\, G \,\, | \,\, gf = g \rbrace.$
Because $g$ acts by isometry, each $g \,\, \epsilon \,\, G$ defines
an element of $SO(2d)$ at a fixed point ($SO(2d)$ may be replaced by some
subgroup depending on the properties which $G$ preserves).
The tangent space for
the orbifold (denoted $T^{\prime}$) at $f$ is thus
\eqn\torbf{T_{f}^{\prime} = {\bf R}^{2d}/S(f).}
On the fixed point sets, i.e. where $S(f)$ is non-trivial, the tangent
space is not a vector space but the cone \torbf,
so the orbifold is not a smooth manifold; it has
a conical singularity.

\subsec{Observables in the Orbifold Sigma Model}

We have already discussed the isomorphism between local operators
(BRST observables) and the cohomology classes of the target manifold.
What, then, are the observables of the topological theory on the orbifold?
 To answer this question, we may begin by recalling the standard lore
or orbifold theories \rDHVW.
 For these theories,
the Hilbert space of the theory splits into a direct sum of
twisted sectors, one for each conjugacy class $\lbrace g \rbrace$ in
the group $G$:
\eqn\hilb{ {\hbox{$\cal H$}} = \bigoplus_{\lbrace g \rbrace}
{\hbox{$\cal H$}}_{\lbrace g \rbrace}.}

In each of these sectors, only the $G$-invariant states
survive the group projection.
A brief word on our notation is in order.  Really, the Hilbert space
splits into one sector per group element.  However, the action of
group elements not commuting with $g$ permutes the sectors in the
conjugacy class of $g.$\footnote{$^1$}{For a string obeying $X(2\pi) =
gX(0)$ we see that $hX(2\pi) = hgX(0) = (hgh^{-1})hX(0).$}  We thus define
\eqn\name{ {\hbox{$\cal H$}}_{\lbrace g \rbrace} =
\bigoplus_{g \, \epsilon \, \lbrace g \rbrace}
{\hbox{$\cal H$}}_{g} =
\bigoplus_{i = 1}
^{| \lbrace g \rbrace |}
{\hbox{$\cal H$}}_{r_i g r_{i}^{-1}} }
for an appropriate set $\lbrace r_i \rbrace.$  We further take the
projection onto $\lbrace g \rbrace$-invariant states.  Now let
\eqn\newhilb{\zeta_{\lbrace g \rbrace} = (\zeta_g,
\zeta_{r_{1}gr_{1}^{-1}},...)
\in {\hbox{$\cal H$}}_{\lbrace g \rbrace}.}
The action of any $h$ in the centralizer, $h \in C(g) \equiv
\lbrace k | kg = gk \rbrace,$ on
$\zeta_{\lbrace g \rbrace}$  is defined by
\eqn\newact{h\zeta_{\lbrace g \rbrace} = (h\zeta_{g}, r_1hr_{1}^{-1}
\zeta_{r_1 g r_{1}^{-1}}, ...).}  This is still
$\lbrace g \rbrace$-invariant.
With these definitions, each ${\hbox{$\cal H$}}_{\lbrace g \rbrace}$ is
invariant as a vector space under the action of any group element.  Thus,
the concept of $G$-invariant states now makes sense, and the state
${1 \over |C(G)|}\sum_{h \, \epsilon \, C(g)} h \zeta_{\lbrace g\rbrace}$
is group invariant.  In effect, we only have to take a $C(g)$ projection.

A similar description of the observables is found for the orbifold sigma
model. Once again we will make notation simpler by eliminating the
conjugacy class label, and only considering $C(g)$-invariant states.  By
the above procedure, in which $g$ represents $\lbrace g \rbrace,$ this
suffices.

As always, we begin with the untwisted sector.  Here we have all
of the observables in the original theory (the cohomology classes of
$K$), and must project onto those which are $G$-invariant.  That is, we
are interested in the differential forms $A$ obeying $g^{*}A =
A.$  Let ${\hbox{$\cal O$}}_1^{p,q}$ represent the untwisted
observables in the orbifold theory with fermion-anti-fermion number
$(p,q),$ where for simplicity in the following we have chosen
the anti-chiral fermion number to be positive; thus the total
fermion number is $p - q.$  (Although the chiral fermion numbers
will only be conserved for Calabi-Yau orbifods, we will be able to make
sense of chiral fermion number violation as we did following \cpnring.)
We see that we have
\eqn\orbobs{{\hbox{$\cal O$}}_1^{p,q} = H_{G}^{p,q}(K),}
where the subscript represents $G$-invariance.  By considering the
Poincar\'e duals of these forms, we may think of them as lying on the
quotient $K/G.$  In this way, we are able to see the equivalence between
$H_{G}^{*}(K)$ and the simplicial cohomology $H_{simp}^{*}(K/G)$ of the
coset space, which is well-defined even though $K/G$ is not a
smooth manifold.  This interpretation allows us to show the familiar
equivalence \rDHVW\ between the untwisted $\Tr(-1)^F$ and the Euler
characteristic.  Since (anti-)chiral fermion numbers correspond to
(anti-)holomorphic form degree, we have
\eqn\wittind{\Tr_{{\hbox{$\cal H$}}_1}(-1)^F = \sum_{p,q = 0}^{d}
h_{G}^{p,q} = \chi_{simp}(K/G).}
In the above formula, ${\hbox{$\cal H$}}_1$ represents the untwisted,
$G$-invariant Hilbert space, and the $h_{G}^{p,q}$ are the Betti numbers
of the $G$-invariant simplicial cohomology.  In fact the value of
\wittind\ may be calculated by considering the operator which projects to
group invariant states, $P = {1 \over |G|}\sum{g}$ (note $G = C(1)$).
Now in the calculation of
$\Tr_{{\hbox{$\cal H$}}_1}(-1)^F$ by the path integral, the presence of
$g$ in the trace yields the Lefschetz number of $g,$ which is the Euler
number of the fixed point sets \rWSYMM.  Hence \rDHVW,
\eqn\morewit{\Tr_{{\hbox{$\cal H$}}_1}(-1)^F = {1 \over
|G|}\sum{\chi(M_g)}.}
Note that \morewit\ agrees
with the right hand side of \wittind, as it
should.\footnote{$^1$}{This was proved, for example in \rRONE.
The basic point is that we may take a simplicial
decomposition of $K$ on which $G$ has a well-defined linear action on
simplices of a given dimension.  Then the
simplices fixed under $G$ form a decomposition of the fixed manifold.
Now, when we sum over the group elements and take the trace, we get zero
from the other simplices and the cardinality of the group for each fixed
simplex.}

Consider now a $g-$twisted ground state, which corresponds to a string
sitting at a point.  If this state is twisted, the point must lie in the
fixed point set of $g.$  Let us call this manifold
$M_g.$\footnote{$^2$}{To see that this space is a manifold,
consider the linear $g$-action on the tangent space of $K$ at a fixed
point $f$ of $g.$  We denote this (push-forward) map by $g_{*}.$
Then the exponential map $\hbox{exp}: T_{f}K \rightarrow K$ will
diffeomorphically map the linear subspace
annihilated by $dg$ onto the fixed point set of $g.$  (Since $G$
respects the metric and complex structure on a Kahler manifold, it
commutes with the connection, and hence the exponential map as well.)
This coordinatization shows why $M_g$ is a manifold.
Similar considerations reduce other
questions about $M_g$ to linear algebra.}
These manifolds will play a crucial role in our analysis, so
we pause here to consider the geometry of these spaces.
It is important for us
to show the complex structure of $M_g.$  In fact, we may use the same $J$
that we used for $K,$ considering the tangent vectors on $TM_g$ as
vectors in the larger space $TK$ (specifically, we use the push-forward
under the inclusion map).
Let $v \,\, \epsilon \,\, T_{f}M_{g}.$ We may express
$v$ as the ``time" derivative of a path $Q(\tau)$ on $M_g,$
i.e. $v = \dot{Q}(\tau_0).$
Now since the action of $g$ is compatible with $J,$ by \comp\ we have
\eqn\tant{g_{*} \circ J(v) = J \circ g_{*}(v) = J \circ
g_{*}(\dot{Q}(\tau_0)) = J(\dot{Q}(\tau_0)) = J(v),}
where we have used the fact that $g_{*}(\dot{Q}) = \dot{Q},$
since $Q$ lies entirely
in the fixed
point manifold $M_g$.  So we see that $J(v)$ is fixed under $g_{*}.$  But
since
\eqn\bund{TK|_{M_g} = TM_g \oplus NM_g,}
where $NM_g$ is the normal bundle on $M_g$, on which $g_*$ acts
nontrivially, we see that $J(v) \,\, \epsilon \,\, T_{f}M_{g},$
which shows that $J$ is a complex structure on $M_g.$  Therefore, it
makes sense to speak of the Hodge numbers of the fixed point manifolds.
The Dolbeault cohomology classes of these spaces will correspond to
observables in the $K/G$ theory.

Finally, we should consider the nature
of the $G$-action on the normal bundle.\footnote{$^1$}{$G$ acts trivially
on the tangent bundle since it fixes all possible paths in $M_g$ and
hence all vectors.}.  We know that $G$ respects the metric, hence also
the volume form.  In a real basis $\lbrace x^i
\rbrace,$ the volume form is a multiple of
$\eta = dx^1\wedge ...\wedge dx^{2d}.$
Now $g^*\eta = \eta$ means that at a fixed point, the pull-back action of
$g^*$ is represented by a matrix in
$GL(2d,{\bf R})$ ($g$ is invertible) satisfying
\eqn\spec{\bigwedge ^{2d}g^* = 1;}
but since the highest exterior power of a matrix is its determinant, we
find that $g^* \in SL(2d,{\bf R}).$  Note that the
same is true of $g_*$, since $g_{*} = [(g^{*})^{T}]^{-1}$ which can
be easily seen by preservation of $<dz^i,{\partial \over \partial
z^j}> = \delta^i_j.$  The above reasoning
extends to the other
conditions we've placed on $g.$  Since $g$ preserves the complex
structure, which means locally that it doesn't mix $z$'s and
$\overline{z}$'s, we can see that $g^* \in U(d)$ in a
complex basis.  We also
require that $g^*$ preserves the holomorphic top forms,
which restricts the
determinant to be unity, i.e. $g^*, g_* \in SU(d).$
But we already know that $g_*$ decomposes into
the identity on $TM_g$ and a matrix which, by abuse of notation, we call
$g.$  That is, $g_* = 1 \oplus g$ in \bund.
Collecting this data, we have that $g$ is $non-trivial$ and
\eqn\sub{g \in SU(\hbox{codim}_{\bf C}M_g).}
As an immediate corollary, we see that for group actions satisfying the
Calabi-Yau conditions we have imposed, there are no fixed manifolds of
codimension one.  We will need this result.

To understand twisted observables one must first understand twisted
interactions, a subject of the next subsection.  Here we will need the
following result, which is proved in that subsection.  Essentially, twisted
observables are also differential forms, but the only
piece which matters in correlations is the value of the
pull-back onto $M_g$ by the inclusion map.
Of course, for the untwisted case, this characterization is still true,
since $M_1 = K.$  Then, BRST cohomology corresponds to differential
cohomology on $M_g.$ Note, then, that the original form $A$ need not be
closed on $K:$  if $i^*A$ is closed, then the non-closed part of
$A$ on $K$ must contribute zero always.
Hence, we have the twisted version of \orbobs\ for the
$g$-twisted sector:
\eqn\twistobs{{\hbox{$\cal O$}}_g \cong H_{G}(M_g),}
where we have intentionally omitted the Hodge labels $(p,q).$  Once
again, we must keep in mind that each observable represents a full
conjugacy class.  In this case, the different sectors within a conjugacy
class are equivalent since $r: M_g \rightarrow M_{rgr^{-1}}$ is a
holomorphic homeomorphism.
For simplicity in labeling, we have dropped the
conjugacy class symbol.

Recall now that the equivalence of (anti-)holomorphic form degree with
(anti-)chiral fermion number was due to the construction of the
observables with fermionic fields of definite chirality.  Implicit in
the above was that the vacuum had fermion
number equal to zero.  This reasoning breaks down in the twisted sector
because of a shift in the fermion number of the vacuum \rJR \rWW.  Although a
constant bosonic field at a fixed point describes a vacuum, the fermionic
fields, even though at a fixed point, cannot be constant - for to be
twisted they must go from one tangent vector to the $g-$translated
vector, and $g$ acts nontrivially on all fields
corresponding to normal directions.  Thus the
fermionic vacuum corresponds to
a sector with generalized boundary conditions on the ends of the string;
the shift in the chiral fermion number of the vacuum
is a general phenomenon for fermions in one real spatial dimension
obeying nontrivial boundary conditions (Originally, in
\rJR, the fermions were in the presence of instantons).
As we have previously discussed,
$g$ acts nontrivially on the normal bundle, $NM_g,$ and trivially
on the tangent bundle $TM_g.$  Focusing
on the chiral fermions, let us imagine just one chiral fermion in
one spatial dimension obeying generalized boundary conditions.
It was shown in \rWW\ that when the (chiral) fermion number is properly
regularized to account for an infinite spectrum of energies,
the more general boundary condition
$\psi(\sigma + 2\pi) = \hbox{e}^{-2\pi i f}\psi(\sigma)$ leads to
the non-zero result\footnote{$^1$}{The proof is straightforward \rWW:
the fermion number of the vacuum is the integral of the energy density
for all energies less than zero.  This is the filled fermi sea.
We regularize this fermion number by subtracting the total number of
fermionic states in the Hilbert space - a (perhaps infinite) constant -
and inswerting a convergence factor.
We have
$F = C - {1\over2} \lim_{s \rightarrow 0}\int_{-\infty}^{\infty}
\hbox{d}E \rho(E) \hbox{sgn}(E) \exp{-s|E|}.$
We choose the constant $C$ to be
$-{1\over2}$ by requiring the
periodic fermionic vacuum to have zero fermion number.
Using the plane wave solutions
$\psi_{n}(\sigma) =
\exp{i(n - f)\sigma}$ with $E_{n} = (n - f)$
yields the result for the boundary conditions stated above.}
\eqn\fermshift{F = f}
for the twisted fermion vacuum.
We take $F = 0$ for the periodic case, corresponding to the unique
Neveu-Schwartz vacuum (the fermionic fields are periodic after twisting).
This argument extends simply to the
anti-chiral and multiple-fermion cases.  Choosing a basis for the tangent
space so that $g$ is diagonal, we can see that we have a separate shift
for each of the chiral fermions.  If the eigenvalues of $g$ are
$\hbox{e}^{-2\pi i f_j},$ $j = 1...n,$ where
$n = \hbox{codim}_{\bf C}M_g,$ then the chiral fermion number of the
vacuum shifts by
\eqn\fersh{F_g = \sum_{j = 1}^{n}f_j.}
where we always take the $f_j$ to satisfy $0 < f_j < 1.$
The formula \fersh\ looks like
$F_g = (i/2\pi)\Tr(\log g) = (-i/2\pi)\log \det g,$ but is of course
different (for example if $\alpha$ is a primitive third root of unity,
then $\hbox{diag}(\alpha,\alpha,\alpha)$ yields $F_g = 1$ or $2$).
However, for Calabi-Yau orbifolds
we do have $\det g = 1,$ which means that $F_g$ is integral.
This gives us
\eqn\bnds{\matrix{0 < F_g < n \hbox{  for } g \neq 1 & F_g \in Z.}}

Some words are in order about the choice in defining fermion number. We
have chosen the untwisted sector to have $F_1 = 0,$ of course, and
have $0 < f_j < 1$ for nontrivial $g.$  The
reasons for this choice are twofold.
One way to set the
fermion number is through interactions.  Namely, the three-point
functions on the sphere determine the ring structure of the observables
(the chiral-primary ring).  For the vacua, these correlation functions
correspond to twist field calculations.  By requiring the twist fields to
respect fermion grading (in the Calabi-Yau case this is possible), we are
led to unambiguous assignments.  We will encounter an example of this in
section five.  The other way of determining the value of the shift is
to consider the path of a twisted string with no oscillator modes.
In one complex dimension, that
path looks like $X(z) = z^{f_j},$ which is non-singular as $z \rightarrow
0$ for $f_j$
positive, and is minimal for $f_j < 1$
(By this we mean that $X(z) = z^{1 + f_j}$ could be thought of as
the product of a twisted string and a closed untwisted string.)
Finally, note that $F_g$ is now
well-defined and independent of which point on
(the connected component of) $M_g$ we choose to determine it; for $g$ has
a finite order, say $m$, so $f_j = k_j/m$, which is fixed, since it
cannot vary continuously along a component of $M_g$ (other components
will correspond to separate operators with different shifts).

We should also point out that the shift is the same in the anti-chiral
sector: $F_a = F_c,$ which yields the same shift of form degrees.
(Remember, we
chose the anti-fermion number to be +1 for an anti-fermion.)
In the anti-chiral sector, $g$ acts by its complex conjugate
(i.e. $f_j \rightarrow 1-f_j$) but the change in the Hamiltonian
compensates for this difference, yielding the desired result.

\subsec{Twisted Interactions, Observables, and Poincar\'e Duality}

In this section we will prove equation \twistobs\ by carefully considering
interactions of twisted observables.  We will also show that the
interpretation of the observables as cohomology elements does not run
counter to Poincar\'e or Hodge duality, or to the Calabi-Yau
characterization.  Namely, we show that the Hodge diamonds of
Calabi-Yau orbifolds have all the
properties one would require of Calabi-Yau manifolds.  Several examples
of orbifold cohomologies are computed; they agree with the cohomologies
derived from the corresponding resolved manifolds or Landau-Ginzburg
orbifolds, where appropriate.

Let us briefly recall the procedure for computing interactions of
orbifolds by the path integral method \rHV.  Consider a loop $X(\sigma)$
twisted such that $X(\sigma + 2\pi) = gX(\sigma).$  As a map from the
Riemann surface, a configuration corresponding to a $g$-twisted state at
$z = 0$ must satisfy $\Phi(\hbox{e}^{2 \pi i}z) = g\Phi(z).$  Orbifold
configurations involve multivalued maps $\Phi: \Sigma \rightarrow K$ with
proper monodromies around points of insertion of twisted states.  We can
find an equivalent description with single-valued maps by choosing a
cover $\tilde{\Sigma}$ of $\Sigma$ on which $G$ acts by automorphism
(preserves metric, complex structure):  $\Sigma \cong \tilde{\Sigma}/G.$
Now for a $g$-twisted state at $z_0,$ we choose our group action such
that a small loop around $z_0$ (i.e., one not enclosing other points of
insertion) will lift to a line from $\tilde{z}$ to $g\tilde{z},$ say.
For an interaction involving observables twisted by $g_1,...,g_n,$ (with
$\prod_i g_i = 1$ for $\Sigma \cong S^2$ the selection rule, viewing all
states as incoming) at $p_1,...,p_n,$ we consider $\tilde{\Sigma},$ a
$G$-cover of $\Sigma,$ with loops around the $p_i$ lifting to lines with
endpoints separated by the action of $g_i.$  In particular,
continuity of the $G$ action for very small loops means that
the $p_i$ descend from fixed points of $g_i$ on $\tilde{\Sigma}:$
$g_i\tilde{p}_i = \tilde{p}_i.$
Now $\tilde{\Phi}:
\tilde{\Sigma} \rightarrow K$ obeying
\eqn\equmap{\tilde{\Phi}(g \tilde{z}) = g \tilde{\Phi}(\tilde{z})}
is a single-valued map with equivalent information.  That is,
instead of $S(\Phi; \Sigma)$ we consider the same theory on
$\tilde{\Sigma}$ with $\tilde{\Phi}$ and the pull-back metric (under the
projection from the cover),
with the exception that we must divide by $N = |G|,$\footnote{$^1$}
{The cover need not be of order $|G|,$ actually.
The order of the cover can be chosen to be the order
of the group generated by the $g_i$.}
since
we have overcounted the area by the order of the covering.  The genus of
$\tilde{\Sigma}$
can be easily obtained from knowledge of $G$ and the orders of the $g_i$
\rHV (see footnote following (5.9));
finding $\tilde{\Sigma}$ explicitly, however, may be very difficult.
Of course the different thing about orbifold interactions is that each
interaction requires a new $\tilde{\Sigma},$ and the functional integral
will be taken only over equivariant maps, i.e. maps obeying \equmap.

Let us turn now to explaining \twistobs.  Generally, a candidate
observable can be likened to a differential form (not necessarily a
cohomology class) as discussed in section two.  There it was explained
that exact forms should be set to zero, while the interest with
BRST-compatible observables forced us to consider cohomology classes.
This analysis must be reconsidered in the case of twisted observables.
For example, suppose we were to consider a correlation function involving
a $g$-twisted state at a point $p$ on $\Sigma.$  Then by the above, we
would need to consider equivariant maps around $\tilde{p}.$
But the equivariant
condition \equmap, together with our observation that $\tilde{p}$ lies at
a fixed point of $g$ on $\tilde{\Sigma}$ means that at $\tilde{p}$
we have $\tilde{\Phi}(g\tilde{p}) = \tilde{\Phi}(\tilde{p}) =
g\tilde{\Phi}(\tilde{p}).$  Hence $\tilde{\Phi}(\tilde{p}) \,\, \epsilon
\,\, M_g.$  This is an important observation!  For instance, suppose we
consider a differential form $A \neq 0$ on K.  If, however, the
restriction of $A$ to $M_g$ were equal to zero, i.e. $A|_{M_g} = 0,$
then the observable corresponding to $A$ would always be evaluated at
some $\tilde{\Phi}(\tilde{p}) \equiv m \,\, \epsilon \,\, M_g$
and hence always give zero.
Thus {\sl as a }g{\sl -twisted observable,} we could set
${\hbox{$\cal O$}}_A = 0;$ for every correlation
function with ${\hbox{$\cal O$}}_A$ would be zero.

Of course the (not necessarily closed) form $A$ must be invariant under
pull-back by the group action to be
an observable in the orbifold theory.  This gives us a condition for its
value at $m$, which we just saw had to lie in the fixed manifold
$M_g.$  For suppose $A$ had indices pointing in the direction normal to
$M_g.$  We know that $g$ acts $nontrivially,$ i.e. has nonunital
eigenvalues, in these directions, so
invariance at $m$ - which is invariance under the
differential matrix $g$
- is impossible!  Any normal components are projected out.  Thus, we see
\eqn\abuse{A|_{M_g} = i^{*}A + (\hbox{Noninvariant Terms} \rightarrow 0)}
where we have abused notation slightly by considering $i^*A$ as a form on
$TK|_{M_g}$ (this can be done because there is a 1-1 imbedding of $TM_g$
into $TK|_{M_g}.)$  Thus, since all values of $A$ outside the
manifold $M_g$ do not contribute to correlation functions, and since
the normal components of $A|_{M_g}$ are projected out, the observable
$A$ is completely determined by $i^*A.$

We are now ready to classify the $g-$twisted observables.  Since $M_g$ is
imbedded in $K,$ the map $i^*: \Omega^*(K) \rightarrow \Omega^*(M_g)$ is
onto.  Therefore, forms expressable as $i^*A$ are isomorphic to all
differential forms on $M_g.$  Now, since $i^*$ and $h^*$ commute, where
$h \in C(g),$ the
invariant forms are just the $C(g)-$invariant forms on $M_g$ (as
always, $g$ represents any element of the conjugacy class $\lbrace g
\rbrace$).  Finally, we must impose the BRST symmetry, which means only
considering forms such that $di^*A = 0$ modulo all forms $i^*dA$.
But pull-back commutes with exterior
derivative, and so we must take all closed forms and mod out by all
exact forms on $M_g$ (remember
$i^*$ is onto).  Thus we have shown \twistobs.

We now know the observables associated to an orbifolded topological sigma
model.  Furthermore, by carefully keeping track of the fermion number
shift associated to
twisted boundary conditions on fermions, we were able to assign the
correct fermion numbers to these observables.  Now by analogy with the
untwisted case, it is tempting to assert that these observables
correspond to cohomology classes associated to the singular space $K/G,$
with holomorphic form degrees given by the chiral fermion numbers.
So let us consider what the Hodge diamond of such a space would
be.  In several examples, we show agreement with the Betti numbers of the
resolutions of orbifolds.  It is not clear whether the two models would
yield the same physics.

In the $g$-twisted sector, the vacuum has chiral-anti-chiral fermion
number $(F_g,F_g),$ as we've defined it.
We saw in \twistobs\ that the space of observables
in this sector was isomorphic to the $C(g)$-invariant cohomology of
$M_g.$  These observables are built from untwisted fermion
operators, which have their usual fermion numbers (form degrees), and
the twist field part which shifts the vacuum.  For example, the identity
operator in the twisted sector is the actual
twist field.  Thus these operators have their degrees
shifted by $(F_g,F_g).$  We may thus {\sl define}
the twisted Hodge numbers of the orbifold $K/G$ to be given by
\eqn\twistcohom{H^{p,q}(K/G) \equiv \bigoplus_{\lbrace g \rbrace}
H^{p-F_g,q-F_g}_{C(g)}(M_g)}
for any $g$ representing $\lbrace g \rbrace.$

Does this definition preserve the familiar structure of the Calabi-Yau
Hodge diamond?  Yes.  This is easily seen by realizing that $M_g =
M_{g^{-1}}$ and $C(g) = C(g^{-1}).$
Thus if $\lbrace k_j/m \rbrace$ represents the $f_j$ for
the action of $g$ on the normal bundle,
then $\lbrace (1 - k_{j}/m \rbrace$ represents
$g^{-1}.$  We conclude that
\eqn\fff{F_{g^{-1}} = n - F_g,}
where $n = \hbox{codim}_{\bf C}M_g.$
This shows us that
there is indeed Poincar\'e duality.  Namely, let $\theta$ be a
$C(g)$-invariant $(p,q)$ form on $M_g.$  Then in the language of
\twistcohom:
\eqn\formex{\theta \in H^{p+F_g,q+F_g}(K/G); \,\,
0 \leq p,q \leq n}
Let $\tilde{\theta}$ be the Poincar\'e dual of $\theta$ in $M_{g^{-1}} =
M_g.$  As observables, the Poincar\'e dual of $\theta$ is $\tilde{\theta}$
in the sector twisted by $g^{-1}.$  This is easily seen; for
\eqn\dualex{\tilde{\theta} \in
H^{(\hbox{dim}_{\bf C}M_{g^{-1}} - p+F_{g^{-1}},
\hbox{dim}_{\bf C}M_{g^{-1}} - q+F_{g^{-1}})}(K/G).}
Using \fff, we see
\eqn\pdualone{\hbox{dim}_{\bf C}M_{g^{-1}} - p+F_{g^{-1}} = d - (p+F_g)}
and likewise for $q.$  So Poincar\'e duality of $K/G$ is shown.
We note here that
$\tilde{\theta}$ has the product structure of a Poincar\'e dual as
well.  That is, if we consider the correlation function
$\langle\theta(p)\tilde{\theta}(q)\rangle$ on the sphere, then by considering
$\theta$ and $\tilde{\theta}$ to have support only on their Poincar\'e
duals, these will intersect only at a single point, call it $x$.
So, going to the
$N$-fold cover of the sphere with two fixed points, where $N$ is the
order of $g$ (i.e. another sphere,
with $g$ acting as rotation by $2\pi/N$), we find a single
equivariant holomorphic
map of degree zero - the constant map $x.$  Thus
\eqn\pdual{\langle\theta(p)\tilde{\theta}(q)\rangle = 1.}
This suggests $\theta \tilde{\theta} = X,$ where $X$ represents the
volume form on $K$ (which is an untwisted observable).  As we have
discussed, however, nonabelian observables are composite operators.
This complicates the product structure.  The identity \pdual\
requires the knowledge of the dimension of moduli space of
equivariant holomorphic
maps of given degree; we must show that there is no higher component of
moduli space of dimension $d$ containing an equivariant map with this
property.  This is easily seen in the examples we compute, as the
dimension increases with instanton number.

We must also show that the $(p,0)$ and $(0,q)$ cohomological structure
of a Calabi-Yau manifold is
preserved in the $K/G$ theory.
In fact the above proof suffices to show this.  Since $F_g >
0$ for all non-trivial $g,$ we see from \twistcohom\ that no twisted
sector can contribute to $H^{*,0}(K/G)$ or $H^{0,*}(K/G).$  By the
duality proven above, the same is true for $H^{*,d}(K/G)$ and
$H^{d,*}(K/G).$  Finally, since the volume and
holomorphic top forms are group
invariant, the familiar structure of the Hodge diamond for Calabi-Yau
orbifolds is preserved.  As an example, let $K$ be a three-fold with $M_g$
codimension two (codimension one is impossible by \sub).  Now \bnds\ tells
us that $F_g = 1$ and the Hodge diamond of $M_g$ fits right in the center
of the diamond $H^{**}(K/G).$

To what extent can we show the equivalence of our cohomology with the
standard cohomology of the resolved manifold?  We know of no general
proof (nor is there a complete understanding of the relationship between
the Landau-Ginzburg models and geometry - see \rRONE \rRTWO \rBGH
and references
therein).  Let us thus concentrate
on a less lofty equivalence - that of the Witten index or Euler
characteristic.  As was shown quite generally in \rDHVW, the Witten index
can be computed for the orbifold theory to be
\eqn\moeuler{\chi_{orbifold}(K/G) = \sum_{\lbrace g
\rbrace}{1\over|G|}\sum_{h \,\, \epsilon \,\, C(g)}\chi(M_{g,h})}
Now by the footnote below \morewit, we can see that we are simply
computing the Euler number of $C(g)$ invariant forms for each $M_g.$
Thus,
\eqn\indagain{\chi_{orbifold}(K/G) = \sum_{\lbrace g
\rbrace}\chi_{C(g)}(M_g).}
Since the fermion number shift of the observables always changes the form
degree by $2F_g$, an {\sl even} number for a Calabi-Yau orbifold,
the ``Euler number" of our orbifold,
calculated directly from the counting of observables,
agrees with expectations.

\newsec{Hodge Numbers of Orbifolds:  Some Examples}

As a concrete example of a Calabi-Yau orbifold which is not
expressible as a complete intersection (and hence has no simple Landau
Ginzburg description - see \rGVW)
we may consider the $Z$ orbifold
$Z = (T \times T \times T)/{\bf Z}_3,$ where $T$ is a torus with
modular parameter $\tau = \hbox{e}^{2\pi i/3}$ and the ${\bf Z}_3$
group action is gemerated by diagonal multiplication by $\tau.$
Note that in this example $T \times T \times T$ is not Calabi-Yau,
but the quotient gives a group-invariant cohomology with Calabi-Yau
structure.  There are $27$ fixed points of this action, all of which
have a $Z_3$ action on the (three-dimensional) normal bundle which is
simply diagonal multiplication by $\alpha^{-1}$ (we must remember that
tangent vectors transform contravariantly).
Now let $g$ be the generator of the ${\bf Z}_3$ action.  We
have $(f_1,f_2,f_3) = (1/3,1/3,1/3),$ which gives us $F_g = 1.$  Thus,
the $27$ vacua in the $g$-twisted sectors all contribute to $H^{1,1}(Z).$
 In the $g^2$ sector, $F_{g^2} = 2,$ so we have a contribution of $27$
elements to $H^{2,2}(Z).$  In the untwisted sector, the invariant forms
contribute nine elements ($dz_id\overline{z}_j$) to $H^{1,1}(Z)$ and
also nine (the duals) to $H^{2,2}(Z),$ in addition to the standard volume,
identity, holomorphic and anti-holomorphic forms.  This analysis agrees
with the Hodge structure of the resolution of $Z$ \rC.

Let us now compute another example which can
be directly compared to a resolved
manifold.  We start with the quintic hypersurface $K$
in ${\bf CP}^4$ defined
by the zero locus of the homogeneous polynomial $W(X) = \sum_{i =
1}^{5}X_{i}^5.$  Now automorphisms of ${\bf CP}^4$
(given by $PGL(5)$) which leave $W(X)$ fixed
will act on $K.$  Let us consider the orbifold of $K$ by
$G = {\bf Z}_5,$ where the generator $g$ of $G$ acts by
\eqn\grpact{g: \hbox{    }
(X_1,X_2,X_3,X_4,X_5) \rightarrow
(X_1,\alpha X_2,\alpha X_3,{\alpha}^4 X_4,{\alpha}^4 X_5),}
where $\alpha = \hbox{e}^{2\pi i/5}.$
If we recall \rCHSW\ that the holomorphic three-form
has a polynomial representation as $\prod_{i} X_{i}^{3}$
then we can easily
see that this form is preserved (because the transformation acts
analytically, $G$ respects the complex structure as well).

Now it is
simple to do our fixed point analysis.  First, in the untwisted sector,
we search for invariant forms.  $K$ has Hodge numbers $h^{1,1} = 1,
h^{2,1} = 101.$  The Kahler form (equivalent to complex structure) is
preserved since $G$ acts holomorphically, so it remains to calculate
which of the $101$ forms of $H^{2,1}$ are invariant.  These have
representatives as homogeneous polynomials of degree five, modulo the
polynomial ideal generated by $X_i^4$ \rCHSW.
Those monomials which remain invariant under the action
\grpact represent invariant forms \rWWW.  It is not difficult to see that
\eqn\invpoly{X_1^3AU \,\,(4), X_1 A^2 U^2 \,\,(4),
A^3B^2 \,\, (4), U^3V^2 \,\, (4), X_1ABUV \,\, (1)}
represent the seventeen invariant  forms in $H_G^{2,1},$
where $A \neq B$ range over $X_2,X_3$ and $U \neq V$ range over $X_3,X_4$
(the numbers of such forms are in parentheses).  First note that $g$
has ten fixed points:
\eqn\fps{M_g = \lbrace (0, 1, -\alpha^m, 0, 0), (0,0,0,1, -\alpha^n): n =
0,...,4 \rbrace}
(this is the full set of fixed points - others are related by projective
equivalence).
Note that all group elements have the same fixed point sets; since these
are discrete, we see that the vector bundle $NM_g,$ which is trivial, has
rank three.  What is the action of $g$? Let us consider the point $p,$
an element of the first set of
five fixed points listed in \fps.  We can coordinatize the manifold $K$
near $p$ by $(\epsilon_1, 1 + \epsilon_2, -\alpha^m + \epsilon_3,
\epsilon_4,\epsilon_5).$  Now we may fix $\epsilon_2 = 0$ by projective
invariance, and use the defining quintic equation for $K$ to determine
$\epsilon_3.$  In this way $(\epsilon_1,\epsilon_4,\epsilon_5)$ represent
a basis for differentials near $p.$  It is simple to see then that $g$
acts by $\hbox{diag}(\alpha^4,\alpha^3,\alpha^3)$ on these differentials.
Since tangent vectors transform contravariantly to differentials, we find
that $(f_1,f_2,f_3) = (4/5,3/5,3/5)$ (recall the hidden $(-)$sign) and
thus $F_g = 2.$  Analyzing the second set of fixed points for $g$ gives
$F_g = 1.$  Thus, in the $g$-twisted sector, we have
$h_g^{2,2} = h_g^{1,1} = 5,$ since each fixed point has a single invariant
cohomology element.  The same is true for $g^2, g^3, g^4.$  We conclude
that the orbifold observables have the structure exhibited in the figure
below:  $h^{1,1}(K/G) = h^{2,2}(K/G) = 21;
h^{2,1}(K/G) = h^{1,2}(K/G) = 17.$
$$\hbox{  }$$

\vfill
$\hbox{                   }\,\,\, \,\hbox{    }
\,\, \hbox{      } \,\,\, \,\,\, h_1^{**}$
\vbox{\tabskip=0pt \offinterlineskip
\def\tablerule{\noalign{\hrule}}
\halign to 1.7in{\strut#& \vrule height25pt#\tabskip=2em minus2em&
\hfil#\hfil &\vrule# &\hfil#\hfil &\vrule# &\hfil#\hfil &\vrule#
&\hfil#\hfil & \vrule#\tabskip=0pt\cr
\tablerule
&& $1$ && $0$ && $0$ && $1$ &\cr\tablerule
&& $0$ && $17$ && $1$ && $0$ &\cr\tablerule
&& $0$ && $1$ && $17$ && $0$ &\cr\tablerule
&& $1$ && $0$ && $0$ && $1$ &\cr\tablerule
}}

$$\hbox{  }$$

$\hbox{       }\sum_{g \neq 1}h_g^{**}$
\vbox{\tabskip=0pt \offinterlineskip
\def\tablerule{\noalign{\hrule}}
\halign to 1.7in{\strut#& \vrule height25pt#\tabskip=2em minus2em&
\hfil#\hfil &\vrule# &\hfil#\hfil &\vrule# &\hfil#\hfil &\vrule#
&\hfil#\hfil & \vrule#\tabskip=0pt\cr
\tablerule
&& $0$ && $0$ && $0$ && $0$ &\cr\tablerule
&& $0$ && $0$ && $20$ && $0$ &\cr\tablerule
&& $0$ && $20$ && $0$ && $0$ &\cr\tablerule
&& $0$ && $0$ && $0$ && $0$ &\cr\tablerule
}}

$$\hbox{  }$$

$\hbox{       }h^{**}(K/G)$
\vbox{\tabskip=0pt \offinterlineskip
\def\tablerule{\noalign{\hrule}}
\halign to 1.7in{\strut#& \vrule height25pt#\tabskip=2em minus2em&
\hfil#\hfil &\vrule# &\hfil#\hfil &\vrule# &\hfil#\hfil &\vrule#
&\hfil#\hfil & \vrule#\tabskip=0pt\cr
\tablerule
&& $1$ && $0$ && $0$ && $1$ &\cr\tablerule
&& $0$ && $17$ && $21$ && $0$ &\cr\tablerule
&& $0$ && $21$ && $17$ && $0$ &\cr\tablerule
&& $1$ && $0$ && $0$ && $1$ &\cr\tablerule
}}

$$\hbox{  }$$
Indeed the above numbers agree with the Hodge numbers of the resolved
manifold of this singular space \rGP.  The same result can be obtained
by considering an apropriate Landau-Ginzburg orbifold \rGVW \rIV \rVMPL.
Namely, the
topological sigma model on $K$ corresponds to the $N = 2$
superconformal Landau-Ginzburg model with superpotential $W = \sum_{i =
1}^{5}\Phi_i^5,$ orbifolded by the group $j,$ which is generated by $j =
\hbox{e}^{2\pi i J_0}$ \rGVW.  If we consider the orbifold of this theory
(i.e. we take $W/(j \times G)$), then a careful treatment of the $U(1)$
charges leads to this theory:  one must identify the Hodge numbers
$(p,q)$ with $(J,3-\overline{J})$ of the NS sector of the $N = 2$ LG
theory.
Note that we are only interested in $j$ cosets; for example,
we consider all elements in
the $g,gj,gj^2,gj^3,gj^4$ sectors to lie in the $g$-twisted sector, and
of course only consider group invariant states \rIV.

Let us consider a case involving a fixed manifold.
Again we consider $W(X) = \sum_{i = 1}^{5}X_{i} = 0$ in
${\bf CP}^{4}.$  We now orbifold by the
${\bf Z}_5$ group generated by $g$:
\eqn\orbact{g: \,\, \hbox{  }(X_1,X_2,X_3,X_4,X_5,)
\rightarrow (\alpha X_1,\alpha^{4} X_2,X_3,X_4,X_5,).}
Now $M_g = \lbrace {\bf X} \, \epsilon \, K: X_1 = X_2 = 0 \rbrace.$
This is clearly a one (complex) dimensional space; we can compute its
Euler number by a simple application of the adjunction formula for Chern
classes (see, e.g., \rGVW).  Since $M_g$ is defined by the zero locus of
the three polynomials
$W, X_1, X_2$ of orders $5, 1, 1,$ we have
\eqn\adjunct{c(M_g) = { (1 + J)^5 \over (1 + 5J)(1 + J)(1 + J)} = 1 - 2J}
where $J$ is the Kahler form.  This yields $\chi(M_g) = -10.$  Of
course all forms are group-invariant since they are invariant under
$g,$ the generator (this is true in all twisted sectors since the order
of the group is prime).We may now
use $(\epsilon_1, \epsilon_2)$ as infinitesimal coordinates normal to
$M_g.$  Then $g^*$ acts by $\hbox{diag}(\alpha,\alpha^4),$ which gives $F_g
= 1,$ as it must for the Hodge diamond of $M_g$ to fit into the orbifold
cohomology without disturbing the Calabi-Yau properties.  This same
structure is repeated for each of the four non-trivial group elements.
The results are summarized in the following tables:
$$\hbox{  }$$

\vfill
$\hbox{                   }\,\,\, \,\hbox{    }
\,\, \hbox{      } \,\,\, \,\,\, h_1^{**}$
\vbox{\tabskip=0pt \offinterlineskip
\def\tablerule{\noalign{\hrule}}
\halign to 1.7in{\strut#& \vrule height25pt#\tabskip=2em minus2em&
\hfil#\hfil &\vrule# &\hfil#\hfil &\vrule# &\hfil#\hfil &\vrule#
&\hfil#\hfil & \vrule#\tabskip=0pt\cr
\tablerule
&& $1$ && $0$ && $0$ && $1$ &\cr\tablerule
&& $0$ && $25$ && $1$ && $0$ &\cr\tablerule
&& $0$ && $1$ && $25$ && $0$ &\cr\tablerule
&& $1$ && $0$ && $0$ && $1$ &\cr\tablerule
}}

$$\hbox{  }$$

$\hbox{       }\sum_{g \neq 1}h_g^{**}$
\vbox{\tabskip=0pt \offinterlineskip
\def\tablerule{\noalign{\hrule}}
\halign to 1.7in{\strut#& \vrule height25pt#\tabskip=2em minus2em&
\hfil#\hfil &\vrule# &\hfil#\hfil &\vrule# &\hfil#\hfil &\vrule#
&\hfil#\hfil & \vrule#\tabskip=0pt\cr
\tablerule
&& $0$ && $0$ && $0$ && $0$ &\cr\tablerule
&& $0$ && $24$ && $4$ && $0$ &\cr\tablerule
&& $0$ && $4$ && $24$ && $0$ &\cr\tablerule
&& $0$ && $0$ && $0$ && $0$ &\cr\tablerule
}}

$$\hbox{  }$$

$\hbox{       }h^{**}(K/G)$
\vbox{\tabskip=0pt \offinterlineskip
\def\tablerule{\noalign{\hrule}}
\halign to 1.7in{\strut#& \vrule height25pt#\tabskip=2em minus2em&
\hfil#\hfil &\vrule# &\hfil#\hfil &\vrule# &\hfil#\hfil &\vrule#
&\hfil#\hfil & \vrule#\tabskip=0pt\cr
\tablerule
&& $1$ && $0$ && $0$ && $1$ &\cr\tablerule
&& $0$ && $49$ && $5$ && $0$ &\cr\tablerule
&& $0$ && $5$ && $49$ && $0$ &\cr\tablerule
&& $1$ && $0$ && $0$ && $1$ &\cr\tablerule
}}

$$\hbox{  }$$

Once again, we find complete agreement with the appropriate
orbifold of the corresponding Landau-Ginzburg model.

As a final example of computing the Hodge numbers of an orbifold, we
consider the following mirror pair.  Let
\eqn\AA{W = \sum_{k = 1}^{5}X_k^5 - 5\psi\prod_{k=1}^{5}X_k}
define a variety $M = (W = 0) \subset {\bf CP}^4.$  This is a Calabi-Yau
space, as is easily seen from the adjunction formula.  Note that $W$ is
the most general quintic invariant under the $Z_5\times Z_5\times Z_5$
action generated by diagonal multiplication by
\eqn\gact{\eqalign{g_1 &= (\alpha^1,1,1,1,\alpha^4) \cr
g_2 &= (1,\alpha^1,1,1,\alpha^4) \cr
g_3 &= (1,1,\alpha^1,1,\alpha^4). }}
Note $g_4 = (1,1,1,\alpha^1,\alpha^4) = (g_1g_2g_3)^{-1}.$

We must determine the fixed point structure of each of the 125 elements
of the group.  This is simplified by noting that whenever more than one
homogeneous coordinated is multiplied by the same power of $\alpha,$ then
there will be a fixed point set determined by setting all other
coordinates to zero.  The results are summarized in the following table,
where we have denoted any (complex) curve by $C,$ and a number
indicates the number of discrete fixed points; group elements
are denoted by the exponents of $\alpha:$  e.g.
$g_1 = (1,0,0,0,4).$

\eqn\bigtable{\matrix{
\sharp&g&\hbox{Example}&M_g&\chi&\chi_{\hbox{inv}}\cr
---&-----&-----&---&---&--- \cr
1&1&(0,0,0,0,0)&M&-200&0 \cr
12&g_i^n&(1,0,0,0,4)&C&-10&2 \cr
\cr
12&\matrix{g_i^{n_i}g_j^{n_j}\cr n_i+n_j=5}&(1,4,0,0,0)&C&-10&2 \cr
\cr
24&\matrix{g_i^{n_i}g_j^{n_j}\cr n_i+n_j\neq5}&(1,2,0,0,2)&10&10&2 \cr
\cr
12&g_i^{n_i}g_j^{n_i}&(1,1,0,0,3)&10&10&2 \cr
\cr
24&\matrix{g_i^{n_i}g_j^{n_j}g_k^{n_k}\cr n_i\neq n_j\neq n_k \neq n_i}&
(1,2,3,0,4)&0&0&0 \cr
\cr
12&\matrix{g_i^{n_i}g_j^{n_j}g_k^{n_k}\cr \sum{n_i} \in 5Z}&(1,1,3,0,0)&
10&10&2 \cr
\cr
12&\matrix{g_i^{n_i}g_j^{n_j}g_k^{n_k}\cr 3{n_i} + n_k \in 5Z}&
(1,1,2,0,1)&C&-10&2 \cr
\cr
12&\matrix{g_i^{n_i}g_j^{n_j}g_k^{n_k}\cr 2{n_i} + 2n_k \in 5Z}&
(1,1,4,0,4)&10&10&2 \cr
\cr
4&(g_1g_2g_3)^n&(1,1,1,0,2)&C&-10&2
}}

This table was calculated using the $G-$index theorem (or Lefschetz
fixed point theorem) to compute
group-invariant cohomology, as described below.
For the elements with isolated fixed points,
the group-invariant cohomology is just the number of orbits under the
action of the other group elements (things are simplified since this is
an abelian orbifold).  As an example, we consider $g_1g_2 =
(1,1,0,0,3).$  The fixed points are $(1,-\alpha^r,0,0,0)$ and
also $(0,0,1,-\alpha^s,0),$ $r,s = 0...4.$  We now concentrate on
the first set of points.
These points are also fixed under $g_3$.  Now under $g_1$ or $g_2$
these points are mapped to $(1,-\alpha^{r-1},0,0,0)$ and
$(1,-\alpha^{r+1},0,0,0)$ respectively.  So the first class contains only
one orbit under the group action.  Similarly for the second.  Thus
$g_1g_2$ has $2$ orbits of fixed points:  $\chi_{\hbox{inv}} = 2.$  The
analysis is similar for other group elements with isolated fixed points.

Now consider a fixed curve.  All are of the form
\eqn\fixcurve{C = \lbrace X^4 + Y^5 + Z^5 = 0 \rbrace \subset
{\bf CP}^2,}
and the adjunction formula tells us that $\chi = -10;$  since this curve
is a complex manifold, $h_{00} = h_{11} = 1,$ $h_{10} = h_{01} = 6.$  We
also know that the volume form and trivial form $1$ are invariant.  So we
have
\eqn\hinv{h_{10, \hbox{inv}} = h_{01, \hbox{inv}} = {2 - \chi_{\hbox{inv}}
\over 2}.}
Now to figure out $\chi_{\hbox{inv}}$ we need to compute the alternating
sum of invariant cohomology elements of various dimensions.  This is much
like the Euler characteristic, except we must insert a projection
operator for $C(g)$ invariance, i.e.
\eqn\chinv{\chi_{C(g)} = \sum_i (-1)^i\hbox{Tr}1_{H_{C(g)}^i} =
{1\over{|C(g)|}}\sum_i (-1)^i \hbox{Tr}g |_{H^i}.}
The above reduces to a sum over fixed points of $g$ (when the fixed
points are isolated \rAB), where we have the formula
\eqn\gindex{\sum_{j=0}^d(-1)^j\hbox{Tr}g|_{H^j} = \sum_{\hbox{fixed
points}}\hbox{sgn}\left({\hbox{det}(1-dg)}\right),}
where $dg$ is the differential $g-$ action on the cotangent space.  In
this example, the fixed point spaces are all one (complex) dimensional,
so $dg$ acts are a rotation by a phase.  In the real sense, we see that
$\hbox{det}(1-dg) = 2-2\hbox{cos}(\theta) \geq 0,$  with equality only
for $g = 1,$ in which case the $g-$index is just the Euler
characteristic, $\chi = -10$ for any fixed curve.  So we only have to
count the number of fixed points for any $g \neq 1.$

Now how does $G$ act on the fixed curve?  Any fixed curve has the form of
\fixcurve, with the action by the group equivalent to the group generated
by the elements $(1,0,0), (0,1,0), (0,0,1),$ where we have used the
same notation as in \bigtable.  One of these elements is
dependent, say $(0,0,1)$, so the action on a fixed curve is by
$Z_5 \times Z_5$ generated by two elements $a$ and $b$ (it is obviously
fixed under the $Z_5$ of the twisting element).  This
group has $25$ elements, $12$ of which are nontrivial and have
fixed points.  They are of the form $(i,0,0), (0,i,0), (i,i,0).$  Each of
these elements has exactly five fixed points.  Thus for any curve we
insert the projection operator onto invariant states to get
\eqn\anycurve{\sum_{j=0}^2(-1)^j \sharp-\hbox{inv}|_{H^j} =
{1\over 25}\sum_{j=0}^2\sum_{m=0}^4\sum_{n=0}^4
(-1)^j\hbox{Tr}a^mb^n|_{H^j} =
{1\over 25}(-10 + 12\cdot5) = 2.}
Thus, for all fixed curves \hinv\ tells us $h_G^{11} = h_G^{00} = 1,
h_G^{10} = h_G^{01} = 0.$

Now to construct the Hodge diamond from the observables, we just need to
shift by the appropriate amount.  There are $101$ elements with fixed
points.  The curves have codimension two and thus have a shift of one,
fitting in the center of the Hodge diamond.  A simple analysis shows that
half of the $80$ fixed point orbits have a shift of one, half by two.
The invariant untwisted elements have no shift.  Thus, the Hodge diamond
is the same as that of the mirror manifold, obtained by resolving this
orbifold \rCDGP:
$$\hbox{  }$$

\vfill
$\hbox{                   }\,\,\, \,\hbox{    }
\,\, \hbox{      } \,\,\, \,\,\, h^{**}(M/G)$
\vbox{\tabskip=0pt \offinterlineskip
\def\tablerule{\noalign{\hrule}}
\halign to 1.7in{\strut#& \vrule height25pt#\tabskip=2em minus2em&
\hfil#\hfil &\vrule# &\hfil#\hfil &\vrule# &\hfil#\hfil &\vrule#
&\hfil#\hfil & \vrule#\tabskip=0pt\cr
\tablerule
&& $1$ && $0$ && $0$ && $1$ &\cr\tablerule
&& $0$ && $101$ && $1$ && $0$ &\cr\tablerule
&& $0$ && $1$ && $101$ && $0$ &\cr\tablerule
&& $1$ && $0$ && $0$ && $1$ &\cr\tablerule
}}

\hbox{$$ $$}
This is the mirror orbifold of $M.$

\newsec{A Dihedral Orbifold}

Now that we know how to compute the ``cohomology'' of the orbifold, we
would like to compute the ring structure as well.  This involves
computations of intersections on the moduli space of equivariant
holomorphic maps
from appropriate branched covers of the Riemann surface, depending on the
interaction under consideration.  In this section, we offer a detailed
compuation of this quantum ring for a nonabelian orbifold.

We wish to
consider an orbifold of ${\bf CP}^1$ by the dihedral group $D_{4},$
the symmetry group of a square.
Recall that ${\bf CP}^1$ is topologically a sphere, and that all the
point groups act naturally on the sphere, since they are subgroups of the
rotation group.  The dihedral group $D_N$ is generated by an order $N$
rotation $\theta$ and a flip $r$, with the relations
\eqn\defdi{r^2 = \theta^N = 1, \hbox{   } r\theta r^{-1} = \theta^{-1}.}
We take the action on ${\bf CP}^1 \cong {\bf C}\cup \infty$ to
be $r(z) = z^{-1}, \theta(z) = \alpha z,$ with
$\alpha = \hbox{e}^{2\pi i/N}.$
Note that for the even dihedral groups there is a non-trivial center
containing the element $\theta^{N/2}.$  In homogeneous coordinates
$(X,Y)$ for ${\bf CP}^1,$ this group has a representation in $PGL(2)$
given by\footnote{$^1$}{We required action by a holomorphic isometry,
hence the group must act as a subgroup of the automorphisms of ${\bf
CP}^1$, i.e. of $PGL(2).$}
\eqn\projrep{r = \left(\matrix{0&1 \cr 1&0}\right), \hbox{  }
\theta = \left(\matrix{1&0 \cr 0&i}\right).}
Note that this is a projective representation - matrices are only defined
modulo nonzero scale factors.

Let us first discuss the fixed point geometry.  Each nontrivial group
element $g$ acts by a rotation of the sphere ${\bf CP}^1,$ and thus
has two fixed points, which we label $A_g$ and $B_g.$  Let us make the
following definitions for $r:$
\eqn\points{\matrix{ {A_r = \left(\matrix{1\cr1}\right)}&
\hbox{ }& {B_r = \left(\matrix{1\cr-1}\right)} }.}
Thus the
cohomology of $M_g$ is just ${\bf C}\oplus{\bf C}.$
Now we know that we can only use $C(g)-$invariant forms.  Consider the
element $r \in \lbrace r \rbrace.$  We have
$C(r) = \lbrace 1,\, r,\, \theta^2,\, r\theta^2 \rbrace,$ and thus
\eqn\rinv{H_{C(r)}(M_r) = 1_{A_r} + 1_{B_r} \Rightarrow \bf{r}}
since the two fixed points are
related under $C(r).$  In this way, we can find all the observables of the
theory.

Although ${\bf CP}^1$ is not a Calabi-Yau manifold,
and thus the chiral
fermion number is not conserved, we can still try to ascribe chiral
fermion numbers to our observables using the methods described in this
paper.  This will then be conserved by assigning a chiral fermion number
$F_\beta = 2$ to the parameter $\beta,$ representing the instanton action
(recall $X^2 = \beta$ is the ring for the ${\bf CP}^1$ model,
which is still true since $X$ remains as an element in the untwisted
sector).  Quite generally, all elements $g$ of order two in a
one-dimensional complex space must have
$F_g = {1\over2},$ since in a neighborhood of a fixed point at $z = 0,$
we have $g(z) = -z,$  or $dg = -1.$  For $\theta$ the action on a local
coordinate at $A_\theta$ gives $d\theta = \hbox{e}^{2\pi i/4}$ and hence
$F_{A_\theta} = {1\over4}.$  Conversely, $F_{B_\theta} = {3\over4}.$
These observations are tallied below.
\eqn\tally{\matrix{\hbox{Observable:  } {\cal O}&\hbox{Sector}&
F_{\cal O} \cr
--------&-----&----- \cr
1&1&0\cr
X&1&1\cr
{\bf r}&\lbrace r \rbrace&{1/2}\cr
{\bf g}&\lbrace r\theta \rbrace&{1/2}\cr
{\bf \theta}_A&\lbrace \theta \rbrace&{1/4} \cr
{\bf \theta}_2&\lbrace \theta^2 \rbrace&{1/2} \cr
{\bf \theta}_B&\lbrace \theta \rbrace&{3/4} }}

Before computing correlation functions, let us anticipate a symmetry of
the chiral ring.  The automorphisms of the group $D_4$ have a
normal subgroup
know as the inner automorphisms, given by conjugation by the various
elements.\footnote{$^1$}{$N \subset G$ is normal if $aN = Na \,
\hbox{ } \forall
a \in G.$  Let $I$ be the inner automorphisms, $i_g \in I$ represents
conjugation by $g.$  Let $\rho$ be an automorphism.
$I$ is normal because $\rho \circ i_g (x) =
(\rho(g))\rho(x)(\rho(g))^{-1} = i_{\rho(g)} \circ \rho(x).$  So $\rho I =
I\rho.$}  The outer automorphisms are those
defined modulo inner automorphims.  Conjugation acts trivially on our
ring elements by construction, but the outer automorphism should
survive in some form in our ring.
 The group of outer automorphisms of $D_4$ is easily seen to be $Z_2$ and
is generated by $\sigma,$ which is determined by its action on $r$ and
$\theta:$  $\sigma(r) = r\theta,$ $\, \sigma(\theta)=\theta.$

In order to derive the chiral ring, we must compute all the three point
functions of the theory.  There is a subtlety, though.  When we write the
observable $\bf{r}$ we really mean a sum of terms related by conjugation.
 In the case of $r,$ for example, we have a nontrivial centralizer which
includes the element $\theta^2,$ relating $A_r$ and $B_r.$  Thus, we have
\eqn\arr{{\bf{r}} = A_r + B_r + A_{r\theta^2} + B_{r\theta^2}.}  In order
to compute a correlation function involving an $r-$twisted operator at
$p$, we
have to choose an appropriate cover $\tilde{\Sigma}$ over $\Sigma.$ (We
will always take $\Sigma$ to be a sphere, since the genus zero
correlation functions determine the ring of observables.)  The different
choices of points $\tilde{p}$ covering $p$ are related by the group
elements and correspond to different twistings in the conjugacy class.
By the way we constructed our operator {\bf r}, our results will be
independent of this choice.  However, to compute correlations involving
{\bf r} we must choose a particular cover.

We begin by considering some simple correlation functions involving two
twist fields.  An explicit
computation will show us how to generalize our procedure for the more
complicated three-point functions.  In genus zero,
the selection rule states that the product of all twists is the identity
(we consider all states as incoming).  Let us compute
$\langle A_\theta(p)B_{\theta^3}(q)X^r\rangle,$ for example;  the
``pre-operators'' in this correlation function are pieces of a
full-fledged observable - they are only defined for a particular choice
of lift.
The first thing we notice is that the
two twisting elements commute. In fact they generate an abelian $Z_4$ subgroup,
which
means that our cover need only be a $Z_4$ cover of the sphere; the other
elements of $G$ will act trivially.  One can compute general $Z_n$
orbifolds of the sphere by a similar calculation \rCV.
Since it is a twice-twisted correlator, we need a
cover of the sphere, branched by $\theta$ and $\theta^3$ over $p$ and
$q.$  Since we can choose an automorphism of the sphere which takes $p$
to the south pole and $q$ to the north pole, we may choose $p$ and $q$ to
be the points $z = 0$ and $z = \infty.$  The covering surface
$\tilde{\Sigma},$ is also a
sphere, and the $Z_4$ acts by rotation.  If $w$ is the coordinate on
$\tilde{\Sigma},$ then $\theta(w) = iw.$  The covering map is $w
\mapsto w^4,$ or in other words $z = w^4,$ so the lifts of a point
$z$ are given by the four points $w = z^{{1\over4}}.$  At $z = 0,$ a
branching point, there is only one $w,$ and we note that a small circle
around the origin lifts to one whose endpoints are separated by the
action of $\theta.$  Now we need to find the equivariant maps from
$\tilde{\Sigma} \cong {\bf CP}^1$ to the target space $K \cong {\bf
CP}^1.$  We know the (compactified) moduli space ${\cal M} \equiv \lbrace
\Phi:  {\bf CP}^1 \rightarrow {\bf CP}^1 | \Phi \hbox{  holomorphic}
\rbrace$ decomposes into maps of degree $k$, with ${\cal M}_k \cong {\bf
CP}^{(1+1)(k+1)-1}.$  We need to find equivariant maps.  Consider the
general degree $k$ holomorphic map given by (see \modcpn)
\eqn\lmap{\Phi:  (X,Y) \mapsto (\sum \phi_{0l} X^{k-l}Y^{l},
\sum \phi_{1_l} X^{k-l}Y^{l}).}
Now $\theta$ acts by
$Y \mapsto iY,$ so recalling that there is an overall scale
ambiguity, we see that $\phi_{lm}$ is equivariant if $m \equiv l + 1
\hbox{ mod } 4$ and $l$ has ranges over a $fixed$ value mod $4.$
The four values of $m \hbox{ mod }4$ represent the four components of
${\cal M}_k,$ which we label ${\cal M}_{k,m}$  For example, we have
${\cal M}_{9,1} = \lbrace (a_1 X^8Y^1 + a_5 X^4Y^5 + Y^9, b_2 X^7Y^2 +
b_6X^3Y^6) \rbrace.$  The astute reader will recognize from the
form of \modcpn\ that equivariance of $\Phi$ means that it commutes with
the projective group action, and so
the different sectors of ${\cal M}_k$ correspond to different spaces of
intertwiners of projective representations of the dihedral group with
various multipliers \rM.  The group action on the space of homogeneous
polynomials of degree $k$ is obtained by the symmetric tensor product of
the representation on $(X,Y).$

We need the maps which take $(1,0) \mapsto (1,0)$ and $(0,1)
\mapsto (0,1).$  The maps will be ill-defined unless there are terms
like $X^kY^0$ and $X^0Y^k.$  So we require
the $X^k$ term to be in the first
coordinate, and the $Y^k$ term in the second.  Thus we must have $l
\equiv 0$ and $k \equiv 1.$  Let us write $k = 4q + 1.$  Counting $a$'s
and $b$'s, we see that $\hbox{dim}{\cal M}_{4q+1,0} = (q+1)+(q+1)-1,$
where we must subtract one for global rescaling of $a$'s and $b$'s.  The
minimum dimension is one, so we must add the observable $X,$ representing
the volume form, to our correlation function in order to get a non-zero
correlation number (i.e, to have finite intersection of the cycles in the
moduli space): $r = 1.$
Since $X$ will require maps from a given point to a
single point in the target space, $X$ is a linear condition on the $a$s
and $b$s, and so defines a cycle of codimension one.  Thus, we see that
there is a unique map of degree $4a+1$ for the correlation function
$\langle A_\theta(p)B_{\theta^3}(q)X^{2a + 1}\rangle.$  Since we need the
three-point function, we take $a = 0.$  Then $k = 1$ and the
equivariant maps are $(aX,bY).$  If we take the point of insertion for
the observable $X$ to be $(1,1),$ say, and we represent the (dual of
the) volume
form by the point $(c,d),$ then the unique map is just $\Phi(X,Y) =
(cX,dY).$  Now since the degree just counts the instanton number,
let $\beta = \hbox{e}^{-A}$ represent the contribution of instanton
number one ($A$ is the area of ${\bf CP}^1$).  We recall again
that we must rescale the action (and the area) by $S \rightarrow S/N$
for an $N-$fold cover.  We find:
\eqn\twopoint{\langle A_\theta B_{\theta^3}X\rangle = \beta^{{1\over4}} }
Although should really only
consider three point functions to define the ring, we
note here that $\langle A_\theta B_{\theta^3}X^{2a + 1}\rangle =
\beta^a\beta^{{1\over4}}$ is consistent with the known relation
$X^2 = \beta.$  From this we can see how some ring relations are derived.
 For example, from the above, with the knowledge that $X$ is the only
observable with $\langle X\rangle = 1,$ we can guess that
\eqn\guess{A_\theta \cdot B_{\theta^3} = \beta^{1\over4},}
although this product could conceivably
contain other untwisted elements like $X$ - further analsis shows it does
not.  Again let us stress that we are deriving these relationships for a
particular lift to $\tilde{\Sigma}.$  The full ring of observables
({\bf r}, etc.) is independent of this choice.

The procedure is similar for the three-point functions.  We briefly
consider the correlation function $\langle A_r(p_1) B_\theta(p_2)
A_{r\theta}(p_3)\rangle.$
One can apply the Riemann-Hurewicz
formula to find the genus of the appropriate covering
space.\footnote{$^1$}{This formula \rHV \rFK gives the
genus $\tilde{g}$ of the covering space in terms of the orders $\nu_i$ of
the twisting elements, and the cardinality $N$ of the group they
generate:  $2 - 2\tilde{g} = N(2 - 2g) - N\sum(1-{1\over\nu_i}).$}  The
cover is once again a sphere, where we take the group action to be the
same as for the target space, namely that of \projrep.  We take the lift
of $p_1$ to be $p_r \equiv (1,1)$ (not $(1,-1)$), with $p_\theta \equiv
(1,0)$ and $p_{r\theta} \equiv (\alpha, 1).$

We begin by considering the equivariant map
\eqn\phij{\phi_l: (X,Y) \mapsto (X^{k-l}Y^l, \epsilon X^lY^{k-l})}
where equivariance under $\theta$ and $r$ (and hence all of $D_{4}$)
requires
\eqn\kmod{k \equiv 2l + 1 \hbox{ mod }4,  \hbox{     } \epsilon = \pm 1.}
The general equivariant map will be a sum of the $\phi_l$ of fixed values
of $\epsilon$ and $(l \hbox{ mod } 4).$  Therefore, there are eight sectors of
equivariant maps of a fixed degree.  As before, the different sectors
have different properties, sending
$p_r, p_\theta,$ and $p_{r\theta}$ to different fixed points in
${\bf CP}^1.$  Some sectors drop out, all maps being multiples by $XY$ of
other maps (of two degrees less), and hence equivalent.  For example, to
compute the correlation function
$\langle B_rA_\theta B_{r\theta}\rangle,$ we find that
there is a unique map of degree one, namely $(X,-Y),$ which gives
\eqn\rtg{\langle B_r A_\theta B_{r\theta}\rangle = \beta^{{1\over8}}.}
Note that the chiral fermion number is always violated mod $2$ in
correlations.  This allows the ring structure to preserve $F$ as long as
we take $\beta$ to have $F_{\beta} = 2,$
as in the untwisted ${\bf CP}^1$ case.
Similarly, one must compute $all$ three-point functions for
pre-operators.  These include
the abelian ones involving $(r)(\theta^2)(r\theta^2),$ which only require
a four-fold cover of the sphere (by a sphere), since the three group
elements only generate a $Z_2\times Z_2$ subgroup.

Once we have solved for the (now commutative)
chiral ring, we try to find an economical way
of presenting it.  It turns out that all the ring relations are generated
by the following:
\eqn\rels{\eqalign{
{\bf r} \cdot ({\bf \theta_A})^2 &= 4\beta^{1\over4}{\bf r} \cr
{\bf r} \cdot {\bf r} &= 4X + 4\beta^{1\over2}
             - 4\beta^{1\over4}({\bf \theta_A})^2 \cr
({\bf \theta_A})^4 &= 2X - 2\beta^{1\over2}
             + 4\beta^{1\over4}({\bf\theta_A})^2 \cr
X {\bf \cdot \theta_A} &=
\beta^{1\over4}({\bf \theta_A})^3 - 3\beta^{1\over2}{\bf \theta_A}.}}
The other observables are exressable in terms of ${\bf r}$ and ${\bf
\theta_A}$ (for example, the right hand side of the last equation is just
$\beta^{1\over4}{\bf \theta_B}$).  In fact, using the second and third
equations in \rels we can eliminate $X,$ and
make the ring ``dimensionless."
We also normalize the variables in a way
which is most suitable to more general even dihedral group orbifolds.  We
define:
\eqn\newdef{\eqalign{
\rho &\equiv {1\over(4\beta^{1\over2})^{1\over2}}{\bf r} \cr
\phi &\equiv {1\over(\beta^{1\over4})^{1\over2}}{\bf \theta_A}}}
In terms of these generators, the ring of observables is defined by
\eqn\ring{\eqalign{\rho \phi^2 &= 4\rho \cr
2\rho^2 &= \phi^4 - 2\phi^2 \cr
\phi^5 &= 6\phi^3 - 8\phi.}}
This ring contains all the information of the topological theory.  We use
it to define higher genus amplitudes through factorization.  Note too
that the single outer automorphism survives as an automorphism of the
$ring$ of observables.  In the variables of \ring\
it has the form
\eqn\outer{\eqalign{\rho &\rightarrow {1\over2}\rho\phi \cr
\phi &\rightarrow \phi.}}
The ring \ring\ is the ring of observables of a topological sigma model
orbifold on the space ${\bf CP}^1.$  This space is not a
Calabi-Yau manifold.  However, as we let the area of the space go to
infinity, the curvature must go to zero, giving us a Ricci-flat manifold
- the plane.  Thus, as in \rCV, the limit $\beta \rightarrow 0$
should give
the chiral primary ring of a conformal field theory.  In order to take
the $\beta \rightarrow 0$ limit, we should use \newdef\ to recover the
$\beta$ dependence of the ring.  In doing so, we easily obtain the
following chiral-primary ring:
\eqn\confring{\eqalign{\rho \phi^2 &= 0 \cr
2\rho^2 &= \phi^4 \cr
\phi^5 &= 0.}}
We may ask whether this ring is familiar.  Is it the ring of a
Landau-Ginzburg model?  The answer is no.  In fact, it is quite easy to
see that no superpotential could give rise to this ring.  However, the
ring \ring contains an interesting subring.  Let us consider the ring
generated by the elements $\rho$ and $\phi^2.$  In terms of these
generators, the last relation in \ring\ becomes dependent on the others.
It only enters as $\phi^6 = 6\phi^4 - 8\phi^2,$ a simple consequence of
the other equations.  Let us define $x \equiv \phi^2, y \equiv \rho$ (do
not confuse $x$ with the observable $X$).  This subring is then described
by the relations
\eqn\subring{\eqalign{x^2 &= 2y^2 + 2x\cr xy &= 4y}.}
Now $this$ ring has a simple Landau-Ginburg description.  It is the same
as the ring derived from the superpotential
\eqn\super{W = {x^3\over3} - 2xy^2 - x^2 + 8y^2.}
The last two terms in \super\ are the $\beta-$dependent perturbations,
which vanish as $\beta \rightarrow 0.$  In this limit, we recover the
superpotential
\eqn\ade{W_{D_4} = {x^3\over3} - 2xy^2,}
which is non other than the superpotential corresponding to $D_4$ in the
$A-D-E$ classification of $N = 2$ minimal models.  So a subring of the
dihedral ${\bf CP}^1$ orbifold
is the same as the ring of the corresponding dihedral
Landau-Ginzburg series!?  There is no obvious connection.  In fact, we
will show in the next section that this relationship is somewhat
general:  the chiral ring of the $D_{2k}$ orbifold
has a subring described by a perturbation of the
$D_{k+2}$ superpotential $W = x^{k+1} + xy^2$ (up to normalization).  It
is a coincidence that $2\cdot2=2+2.$

\newsec{${\bf CP}^1/D_{N}$}

In this section, we will outline the generalization to orbifolds of
${\bf CP}^1$ by an arbitrary dihedral group $D_{N}.$  Let us first
consider the even case $N = 2k.$  The features of the
previous section are quite general, so we will be brief.  The dihedral
group is defined by \defdi.  When $N = 2k,$ there are two ``flip''
conjugacy classes, $\lbrace r \rbrace$ and  $\lbrace r\theta \rbrace,$ as
before.  We also have the trivial class $1$, the central element
$\theta^k,$ and $k-1$ conjugacy classes $\lbrace \theta^i \rbrace,$ $i =
1...(k-1)$ (here $\lbrace \theta^i \rbrace = \lbrace \theta^i,
\theta^{-i} \rbrace).$

Now to determine the ring, we must compute many correlation functions
involving the twists $(r)(\theta^l)(r\theta^l).$  These turn out to be
very similar to the ones we just computed.  The main difference is in the
factors of $\beta$ in the ring coefficients.  However, by $F$
conservation, we can always determine the correct $\beta-$dependence from
the ``dimensionless'' operators $\rho$ and $\phi.$  Once again, these
generate the ring, though the relations between them are a bit more
complicated.

Consider the $l-$twisted sector, by which we mean the conjugacy class of
$\lbrace \theta^l \rbrace.$  There are two observables in this sector,
which we will label $\phi_l$ and $\phi_{2k - l}.$  Here we define
\eqn\obdef{\phi_l \equiv (\beta^{-{l\over4k}})
(A_{\theta^l} + B_{\theta^{-l}}).}
We use the convention $\phi_0 \equiv 2,$ and the abelian result
$(A_\theta)^{2k} = X$ gives us that $\phi_{2k} =
(\beta^{-{1\over2}})2X \equiv 2\chi$ ($\chi$ is the dimensionless
version of $X$).
We also have the generalization of \guess:
\eqn\genguess{A_\theta \cdot B_{\theta^{-1}} = \beta^{1\over2k}.}
This, combined with another abelian result, $A_\theta A_\theta =
A_{\theta^2},$ allows us to compute all products of the $\phi_l$ in terms
of $\phi_1 \equiv \phi.$  The trick is to derive a recursion
relationship for the $\phi_l.$  Note, for example, that
\eqn\phph{\eqalign{\phi \cdot \phi_1 &= \phi_1 \cdot \phi_1 \cr
&= \beta^{-1\over2k}(A_{\theta} + B_{\theta^{-1}})
(A_{\theta} + B_{\theta^{-1}}) \cr
&= \beta^{-1\over2k}(A_{\theta^2} + B_{\theta^{-2}} + 2\beta^{1\over2k}
\cr &= \phi_2 + \phi_0.}}
More generally, we find the following recursion relation among the
$\phi_i:$
\eqn\recrel{\phi \cdot \phi_n = \phi_{n+1} + \phi_{n-1}}
This is a difference relation which can be solved as follows.  First,
assume that $\phi$ acts as a constant (which it is not);
let's call it $A$.
Then, as for a second order
differential equation, we say that $\phi_n \sim t^n,$ solve for $t$ and
impose boundary conditions.  We easily see that we must have
\eqn\teq{t^2 - At + 1 = 0}
which gives
\eqn\tis{t_{\pm} = (A/2) \pm i\sqrt{1 - (A/2)^2}.}
The general solution is $\phi_n = c_+t_+^n + c_-t_-^n.$  We must have
that $\phi_0 = 2$ and $\phi_1 = A.$  This gives
\eqn\fin{\phi_n = t_+^n + t_-^n.}
If we formally put $A = 2\hbox{cos}(z),$
then $t_\pm = \hbox{e}^{\pm iz},$ and we
can easily see that $\phi_n = 2\hbox{cos}(nz).$  The Chebyshev polynomial
$W_n(X)$ is a degree $n$ polynomial in $X$ defined by
(conventions vary)
\eqn\cheb{W_n(X = 2\hbox{cos}(z)) = 2\hbox{cos}(nz).}
We thus have derived
\eqn\ficheb{\phi_n = W_n(\phi).}
Although the recursion relation did not have constant coefficients,
the ultimate justification of this method is that it works!

Now the generalization of \twopoint, along with \obdef, tells us
that
\eqn\morel{\chi\phi_l = \phi_{2k-l}.}
Of course $\chi = (1/2)\phi_{2k},$ so,
in particualar
\eqn\morrel{\phi W_{2k}(\phi) = 2W_{2k-1}(\phi).}
In fact this relation generates all of the equations in \morel.  For
example,
\eqn\morelder{\eqalign{\chi\phi_2 &= \chi(\phi^2 - 2) \cr
&= (\chi\phi_1)\phi - \phi_{2k} \cr
&= \phi_{2k-1}\phi - \phi_{2k} \cr
&= \phi_{2k-2},}}
where we have made use of \morrel\ and the recursion relation \recrel.

The ring relations involving $r$ can now be made simpler by defining
\eqn\rodef{\rho \equiv {{\bf r}\over(2k\beta^{1\over2})^{1\over2}},}
where {\bf r} is the conjugacy class operator and contains $2k$ terms,
exactly analogously to \arr.  The simple relations $A_rA_r = X$ and
$A_rB_r = \beta^{1\over2},$ along with their generalizations for the
other flips, are helpful in deriving
\eqn\arrarr{\rho^2 = 1 + \chi + \sum_{l=1}^{k-1}\phi_{2l},}
which we can rewrite as
\eqn\arrsq{\rho^2 = 1 + kW_{2k}(\phi) + \sum_{l=1}^{k-1}2lW_{2l}(\phi).}
Finally, the first relation in \ring survives unchanged with our present
definitions.  This relation exactly parallels the multiplication
of conjugacy classes in the group ring.  Summarizing, the general ring
of observables for the topological orbifold ${\bf CP}^1/D_{2k}$ is:
\eqn\genring{\eqalign{\rho\phi^2 &= 4\rho \cr
\rho^2 &= 1 + {1\over2}W_{2k}(\phi) +\sum_{l=1}^{k-1}W_{2l}(\phi) \cr
\phi W_{2k}(\phi) &= 2W_{2k-1}(\phi). }}

The group outer automorphism survives in the ring as before,
and we have defined our generators so
that \outer\ is valid as written.

Once again, our ring has a subring generated by
$x \equiv \phi^2$ and $y \equiv \rho,$ and
the last relation in \genring becomes redundant.  Note that
$W_{2l}(\phi)$ is a degree $l$ polynomial in $x$ alone, so we can define a
degree $k + 1$ polynomial
$F(x)$ such that the right hand side of the second equation in \genring\
is given by $F^{\prime}(x).$ We can write this subring as the chiral ring
associated to the superpotential
\eqn\newsup{W = F(x) - xy^2 + 4y^2.}
This is a perturbation of the $D_{k+2}$ Landau-Ginzburg potential.
The perturbation involves the Chebyshev polynomials, which have been
shown to be integrable \rDVV \rFI, though we don't know whether this model is
integrable.  This is reminiscent of the ${\bf CP}^1/Z_n$ case, where the ring
was found to be that of a perturbed $A_{2n}$ minimal Landau-Ginzburg model.
(For recent work on the relationship of orbifolds to
Landau-Ginzburg models, see \rCVNEW.)

In the odd case, $N = 2k+1,$ there is perhaps only one subtlety.  In
considering the
covering surface of the sphere for the three point function, one must be
careful in choosing the lift.  For example, the sphere covers the sphere
with the usual action, but if we are considering a $(\rho)(\theta)
(\rho\theta)$ correlation, we should make sure the points
representing $\rho$ and $\rho\theta$ do not lie on the same orbit (or
else they represent the same point on the underlying sphere).  For the
odd orbifolds, there is only one ``flip'' conjugacy class, but there are
two operators associated to it, since the two fixed points are not
related by any element.  Proceeding in much the same way as for the even
case, we find the following ring:
\eqn\oddring{\eqalign{\rho\phi^2 &= 4\rho \cr
\rho^2 &= {1\over2}W_{2k+1}(\phi) +\sum_{l=1}^{k}W_{2l-1}(\phi) \cr
\phi W_{2k+1}(\phi) &= 2W_{2k}(\phi).}}
This ring also has the automorphism
\eqn\outertwo{\eqalign{\rho &\rightarrow {1\over2}\rho\phi \cr
\phi &\rightarrow \phi,}}
though now it corresponds to the geometric symmetry corresponding to a
$\theta$ rotation by $\pi,$ which is not a group element.  No connection
to the $D-$series is evident.

\newsec{${\bf CP}^2/D_4$}

Our techniques allow us to compute higher dimensional orbifolds as well.
In this section, we consider the orbifold ${\bf CP}^2/D_4,$ with the
group generators acting by the matrices
\eqn\cptwoact{r = \left(\matrix{1&0&0 \cr 0&0&1 \cr 0&1&0}\right),
\hbox{  }
\theta = \left(\matrix{1&0&0 \cr 0&i&0 \cr 0&0&-i}\right).}
The reason for considering this orbifold is that as we let the area of
the ${\bf CP}^1$ go to infinity, we can obtain a nonabelian conformal
orbifold theory.  Nonabelian orbifolds have not been heavily studied
(though see \rGATO) and little is known about their twist fields.
To how this limit arises, consider the point
$p = (1,0,0).$  This point is fixed under the entire group $D_4.$
Thus, in the conformal limit $\beta \rightarrow 0,$ the space around $p$
becomes ${\bf C}^2$ and the action of the group is given by the
differential action near $p,$ which is the linear action defined by the
bottom two entries of the matrices in \cptwoact.

There are no subtleties in the computation of the ring for this theory.
The ring contains three observables for each conjugacy class, fifteen
total.  Some group elements have fixed spheres, leaving us with a twisted
volume form $V_g$ as an observable.  This is equal to $X\cdot 1_g.$

The ring of observables is generated by three elements
\eqn\ringens{\eqalign{\chi &\equiv {\beta}^{-{1\over3}}X \cr
\mu &\equiv {1\over2}\beta^{-{1\over6}}{\bf r} \cr
\alpha &\equiv {1\over2}\beta^{-{1\over3}}{\bf A}_{\theta} }}
where ${\bf r}$ represents the composite operator associated to the
nontrivial $0-$form on the fixed sphere of $r,$ and ${\bf A}_{\theta}$
represents the operator corresponding to the fixed point $(1,0,0).$  The
defining relations are
\eqn\cptworing{\eqalign{\alpha^7 &= \alpha \cr
\mu^2 &= \alpha^4 \cr
\mu\alpha^6 &= \mu \cr
\chi^3 &= 1 \cr
\mu\chi &= \mu\alpha^4 \cr
\alpha\chi &= \alpha^5.}}
The group automorphism takes the form
\eqn\cptwoauto{\eqalign{\alpha &\rightarrow \alpha \cr
\mu &\rightarrow \mu\alpha^3 \cr
\chi &\rightarrow \chi}}
in this presentation of the ring.

We leave to further study the consideration of orbfifolds by other
groups and higher dimensional spaces.

\newsec{Methods for  Computing Twist Field Correlations}

Our observables are nothing but twist fields - they create twisted
chiral-primary states in the full non-topological sigma model.  Now with
our knowledge of the ring, we have the $\beta-$dependence of the
theory (which means scale dependence since $\beta = \hbox{e}^{-A}$).
There is
another theory we could have gotten from the original sigma model which
is the complex conjugate theory, obtained by performing the twist of the
$N = 2$ theory so that $anti-$holomorphic maps were the instantons.  The
ring of this theory is obtained by complex conjugation.  Now we can use
recent non-perturbative results \rCVTAF\ for computing the metric
\eqn\topmet{g_{i\overline{j}} = \langle \overline{j} | \ i \rangle}
as a function on coupling constant space.  This is the metric of the
full non-topological sigma model, restricted to the chiral states, and
is closely related to Zamolodchikov's metric \rZ\ (see \rCVTAF\ for a
discussion).  In reference \rCVTAF, the
authors derived differential equations for \topmet.  We will consider
here the scale-dependence of this metric.  The non-trivial input is that
as the area of the ${\bf CP}^1$ goes to infinity the curvature goes to
zero, so there is no curvature anomaly and we have a conformal field
theory.  So we expect good behavior of $g_{i\overline{j}}$ as $\beta
\rightarrow 0.$  As was discussed in \rCV\ and \rCVER, demanding
finiteness in this limit can be enough to specify the exact form of
solution to these equations.

Let us see how this works.  Consider a $Z_n$ orbifold of ${\bf CP}^1,$
as in \rCV.  In order to consider the $\beta$ behavior of the theory, we must
find the
operator corresponding to a perturbation in $\beta.$  Because we
constructed the action from the Kahler form, $X$ is the operator
corresponding to $\beta$ variation.  Actually, $-\hbox{ln}\beta = A$
multiplies the $X$ term, so the operator corresponding to $\beta$ is
properly $C_\beta = -{1\over \beta}X.$
The differential equation for the
metric $g$ is \rCVTAF
\eqn\taf{\partial_{\overline{\beta}}( g \partial_{\beta} g^{-1} ) =
[C_{\beta}, gC_{\beta}^{\dag} g^{-1}].}
The metric $g_{i\overline{j}}$ represents a fusion of topological and
anti-topological (in which the anti-holomorphic maps are instantons)
theories.  The states in these two theories are related by the real
structure matrix:
\eqn\realstr{\langle \overline{j}| = \langle i | M^i_{\overline{j}}.}
The topological metric is $\eta_{ij} = \langle \phi_i \phi_j \rangle.$
 From \realstr\ and the definition \topmet, we see
\eqn\realid{M = \eta^{-1}g.}
The $CPT$ conjugate of $|i\rangle$ is $|\overline{i}\rangle.$  Acting
twice by $CPT$ is the identity, so we see
\eqn\realcond{MM^* = (\eta^{-1}g)(\eta^{-1}g)^* = {\bf 1}.}
For our ${\bf CP}^1/Z_n$ example, we have two observables in each sector,
corresponding to the two fixed points (north and south poles).  The
metric $g$ is block diagonal in each sector, while the metric $\eta$
relates $g-$ and $g^{-1}-$twisted sectors (since it involves no
``out'' states).  There is
a symmetry ($g \rightarrow g^{-1}$ or $z \rightarrow z^{-1}$)
equating the north pole in the $g-$sector to the $g^{-1}-$twisted
south pole.  Consider $g \in Z_n.$  On the $g$ and $g^{-1}$
subspace, with basis $\lbrace 1_{g,a} 1_{g,b}, 1_{g^{-1},a},
1_{g^{-1},b} \rbrace$ we have
\eqn\prereal{\matrix{g = \left(\matrix{a&c&0&0 \cr c^* &b&0&0 \cr
0&0&b&c^* \cr 0&0&c&a}\right) & \eta = \left(\matrix{0&0&1&0\cr
0&0&0&1 \cr 1&0&0&0 \cr 0&1&0&0}\right),}}
where we have used hermiticity and the aforementioned symmetry (note that
$a$ and $b$ are real).
Applying \realcond, we find $ab = 1, c = 0$  ($g=g^{-1} \Rightarrow a = b
=1$).  $g$ depends only on $|\beta|$ \rCV, so we can define
\eqn\change{\matrix{x = 4|\beta|^{1\over2}, & \hbox{} & u(x) =
2\hbox{log }\left(a|\beta|^{{n-2l\over 2n}}\right)}.}
We find from \taf\ that $y$ obeys a special form of the
Painlev\'e III equation:
\eqn\pthree{u^{\prime \prime} = {1\over x}u^{\prime} = 4\hbox{sinh }u.}
Now we must require that
\eqn\goodlim{u \rightarrow r\hbox{log}x + s, \hbox{    } r = 2\left({n -
2l\over n}\right)}
in order for $a$ to be finite at $x = 0.$  It turns out \rPIII\
that restricting
the coefficient on the logarithm in \goodlim\ determines $s$ by the
equation
\eqn\seq{\hbox{e}^{s/2} = {1\over 2^r}{\Gamma \left({{1\over2} - {r\over
4}}\right) \over \Gamma \left({{1\over2} + {r\over 4}}\right)}.}
Resolving the morass, we find
\eqn\alim{a(0)= {\Gamma({l\over n})\over
\Gamma(1 - {l \over n})},}
which we use to derive the proper normalization of the twist fields.

For ${\bf CP}^1/D_4$,
we have
already solved for the ring, so we know what multiplication by $X$ is
(recall from \ficheb\ that $X = \beta^{1\over2}\chi =
{\beta^{1\over2}\over2}\phi_4 =
{\beta^{1\over2}\over2}W_4(\phi)).$  For example, ${\bf r}X =
\beta^{1\over2}{\bf r}$ (${\bf r}$ is given in \newdef), which means that the
matrix $C_{\beta}$ has an
invariant subspace of dimension one.  We easily see that the right hand
side of \taf\ is zero, which, combined with the fact that the metric only
depends on $|\beta|,$ tells us that the normalized operator
${1\over2}{\bf r}$ is
independent of $\beta$ (aside from the normalization arising from
$\langle \overline{1}|1\rangle$).  The same is true for {\bf g} and
${\bf \theta}_2.$  The untwisted observables were discussed in \rCVER.  This
leaves us with ${1\over \sqrt{2}}{\bf \theta}_A$ and
${1\over \sqrt{2}}{\bf \theta}_B,$ where we have chosen a convenient
normalization.  In this subspace, the relevant matrices take the form
\eqn\dfourex{\matrix{
{C_{\beta} = -{1\over \beta} \left(\matrix{0&\beta^{1\over4}\cr
\beta^{3\over 4}&0}\right)},&{\eta = \left(\matrix{0&1\cr 1&0}\right)},&
{g = \left(\matrix{a&c^* \cr c & d}\right)}},}
where $g$ is a general hermitian matrix (with no components outside this
subspace due to the selection rule).  The reality constraint
\realcond\ gives us $c = 0, ad = 1,$ so there is only one real variable,
$a.$  It is now clear that the twist operators reduce to simple $Z_4$ twist
operators.  The reason for this is that the fixed points of $\theta,$ for
example, are fixed by an abelian stabilizer group.  In the large limit,
we are left with two copies of the $Z_4$ orbifold, with operators that
create twisted states in both.

The situation is different for our ${\bf CP}^2/D_4$ orbifold.  In that
case, the point $p \equiv (1,0,0)$ was fixed under the entire nonabelian
group.  Now consider the theory in a neighborhood of $p$ as we take
$\beta \rightarrow 0.$  As discussed in section seven, this will
correspond to a nonabelian orbifold of ${\bf C}^2.$  Consider the
$r-$twisted sector.  We have three operators.  Let ${\bf
\theta}^{\prime}/\sqrt{2}$
represent the fixed point $p$ under $\theta,$  with ${\bf
\theta_A}/\sqrt{2}$ and
${\bf \theta}_B/\sqrt{2}$
the operators associated to the two remaining fixed points (similarly to
\newdef).
In this sub-basis we have
\eqn\cbeta{C_{\beta} = -{1\over
\beta}\left(\matrix{0&\beta^{1\over4}&0 \cr 0&0&\beta^{1\over2} \cr
\beta^{1\over 4}&0&0}\right)}
and the topological metric
\eqn\etmetcptwo{\eta = \left(\matrix{1&0&0 \cr 0&0&1 \cr 0&1&0}\right).}
It is clear that $\eta$ is essentially $\hbox{diag}(1,1,-1),$ which
means, from the reality condition \realstr, that the hermitian matrix
$g$ is just a unitary tranformation of an element in the
complexified group $SO(2,1).$  In general, the equations resulting from
\taf\ using \etmetcptwo\ are quite complicated.  Similar equations were
studied in \rWLH, in the context of Landau-Ginzburg models perturbed away
from criticality.  It is not known whether the requirement of regularity
is enough to fixed the values of the metric (the objects of interest to
us) at the point $\beta = 0.$  However, if we {\sl assume} that the
metric is diagonal, then the reality condition gives $g_{0\overline{0}} =
1,$ and $g_{1\overline{1}}g_{2\overline{2}} = 1,$ so we have one real
parameter.  Then if we define
\eqn\bdredef{\matrix{x = 8|\beta|^{1\over2} & b = 2g_{1\overline{1}}}}
we see that $b(x)$ obeys another special form of the Painlev\'e equation:
\eqn\piiibd{b^{\prime \prime} = {1\over b}(b^{\prime})^2 - {1\over
x}b^{\prime} + {1\over x}b^2 - {1\over b}.}
This is called the Bullough-Dodd equation, and was studied in \rBD.
Requiring regularity of $b$ in the limit $x \rightarrow 0$ again specifies the
boundary conditions.  We find $g_{1\overline{1}}
= \Gamma(3/4)/\Gamma(1/4),$ so we
know that a regular limit exists, though we don't know if our ansatz of a
diagonal metric is valid.  We leave this question to further study.

\newsec{Acknowledgements}

I am grateful to C. Vafa for suggesting this problem, and for
many instructive conversations.  I would also like to thank K.
Intriligator and W. A. Leaf-Herrmann for sharing their ken.

\listrefs

\bye